\documentclass[11pt]{article}
\usepackage[total={6in, 8in}]{geometry}
\usepackage{amsthm,amsmath,amsfonts,amssymb}
\usepackage{graphicx}
\usepackage{bm}
\usepackage{threeparttable}
\usepackage{booktabs,multicol,multirow}
\usepackage{parskip}
\usepackage{authblk}
\RequirePackage[authoryear]{natbib}

\title{A Joint Modeling Approach for Clustering Mixed-Type Multivariate Longitudinal Data: Application to the CHILD Cohort Study}

\author[1]{Zhiwen Tan}
\author[2]{Chang Shen}
\author[3]{Padmaja Subbarao}
\author[4]{Wendy Lou}
\author[1,*]{Zihang Lu}
\affil[1]{Department of Public Health Sciences, Queen’s University, Kingston, ON, Canada.}
\affil[2]{Department of Electrical and Computer Engineering, Queen’s University, Kingston, ON, Canada.}
\affil[3]{Division of Respiratory Medicine, Department of Pediatrics, Hospital for Sick Children  \& University of Toronto, Toronto, ON, Canada.}
\affil[4]{Dalla Lana School of Public Health, University of Toronto, Toronto, ON, Canada.}
\affil[*]{Corresponding author: zihang.lu@queensu.ca}

\date{ }

\begin{document}

\maketitle

\begin{abstract}
In epidemiological and clinical studies, identifying patients' phenotypes based on longitudinal profiles is critical to understanding the disease's developmental patterns.  The current study was motivated by data from a Canadian birth cohort study, the CHILD Cohort Study. Our goal was to use multiple longitudinal respiratory traits to cluster the participants into subgroups with similar longitudinal respiratory profiles in order to identify clinically relevant disease phenotypes. To appropriately account for distinct structures and types of these longitudinal markers, we proposed a novel joint model for clustering mixed-type (continuous, discrete and categorical) multivariate longitudinal data. We also developed a Markov Chain Monte Carlo algorithm to estimate the posterior distribution of model parameters. Analysis of the CHILD Cohort data and simulated data were presented and discussed. Our study demonstrated that the proposed model serves as a useful analytical tool for clustering multivariate mixed-type longitudinal data. We developed an R package \textit{BCClong} to implement the proposed model efficiently.  
\end{abstract}

\section{Introduction}
 In epidemiological and clinical studies, identifying subgroups of patients based on longitudinal profiles is a common research interest to understand the developmental patterns of diseases. Appropriately assigning patients into subgroups that share similar characteristics is a critical first step to understanding the disease heterogeneity and discovering underlying biological mechanisms, leading to personalized medicine and targeted treatment. Statistical methods for clustering a single longitudinal trajectory have been well-developed and widely used in many different medical research areas. However, it is very common these days to encounter situations where several longitudinal markers or responses are collected simultaneously in a study and there is a growing interest to examine how multiple longitudinal characteristics could collectively contribute to disaggregating disease heterogeneity.  \par

\begin{figure}[htbp]  
 \centering
   \includegraphics[width=12cm,height=6cm]{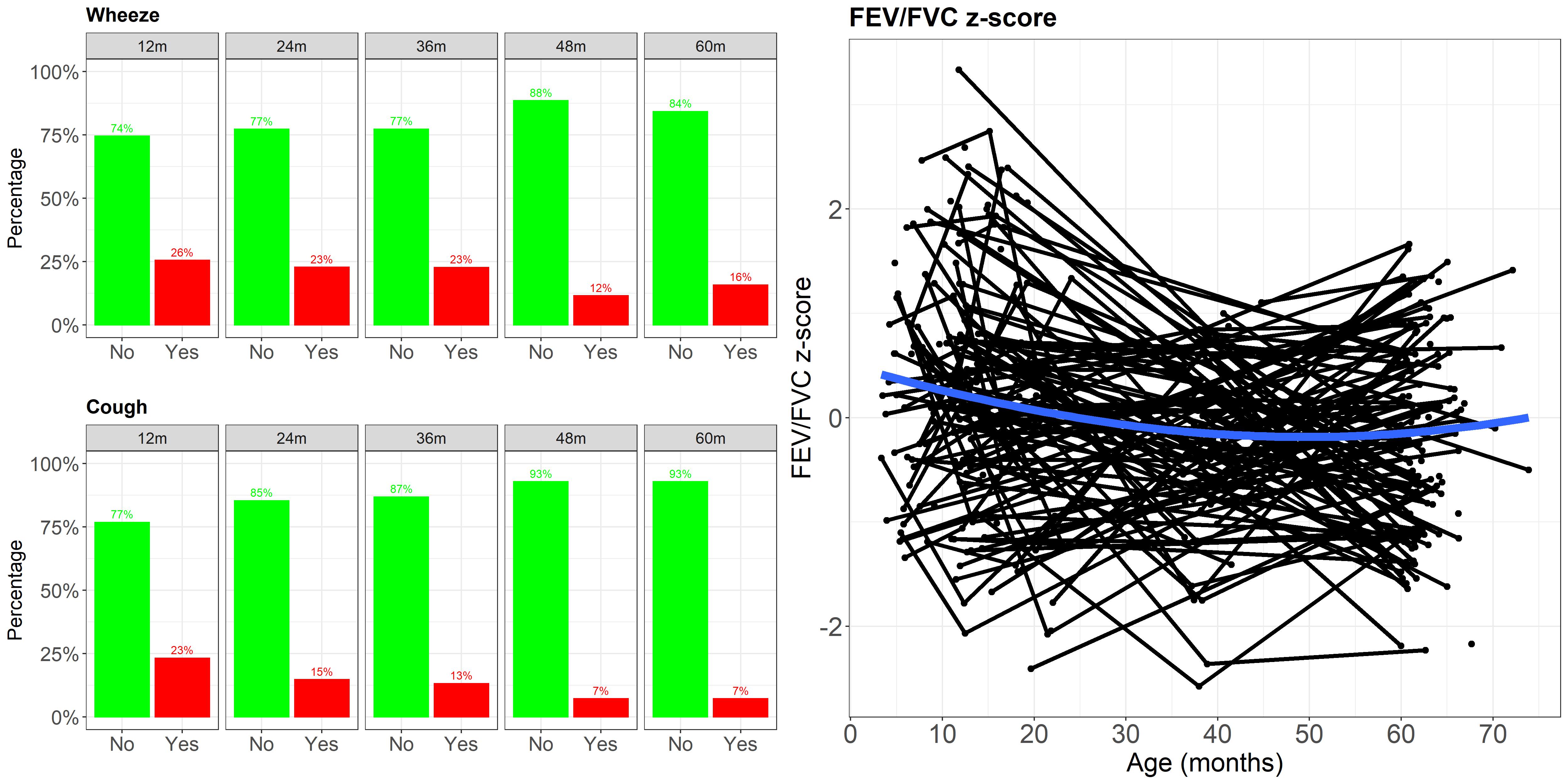}
\caption{Wheeze, cough and lung function (FEV/FVC z-score) distributions over the first 5 years of age in the CHILD Cohort Study.}
\end{figure}

\subsection{Motivating example}
The current study is motivated by data collected from the CHILD Cohort Study \citep{Subbarao2015}. The CHILD Cohort Study is a prospective longitudinal birth cohort study. Pregnant women were initially recruited between 2009 and 2012 in four sites across Canada (Vancouver, Edmonton, Winnipeg and Toronto).  Our goal is to use the longitudinal respiratory traits to cluster the participants into subgroups with similar longitudinal respiratory profiles in order to identify clinically relevant disease phenotypes. This process might reveal distinct disease developmental patterns and discover new phenotypes, which will promote a better understanding of the functional or pathobiological mechanism, leading to targeted interventions and precision medicine. To define these phenotypes, we will focus on the longitudinal variables of wheeze, cough without a cold and lung function (Figure 1). The wheeze and cough data were collected through questionnaires completed by parents at ages 12, 24, 36, 48 and 60 months. The lung function FEV/FVC, measured by the ratio of forced expiratory volume (FEV) and forced vital capacity (FVC), was collected during the scheduled visits at 3, 12, 18, 36 and 60 months. Higher FEV/FVC indicates better lung function.

\subsection{Review of existing statistical methods}
In the statistical literature, numerous models have been developed for clustering longitudinal data under different contexts. When clustering a single longitudinal trajectory, methods that are widely used in practice include group-based latent class analysis \citep{Nagin1999} and latent class mixed effect model \citep{Proust-Lima2015}. The ease of implementing these methods using standard statistical software also makes these methods attractive. For example, the former approach can be carried out using SAS Proc Traj whereas the latter can be carried out using R \textit{lcmm} package. Other methods such as K-means clustering \citep{Genolini2010}, mixture modeling based on a smoothing function \citep{Ding2021} or based on a multivariate $t$ distribution \citep{McNicholas2012} have also been proposed. Longitudinal data can be viewed as a type of functional data and therefore functional clustering methods can also be used for clustering purposes. These methods include model-based functional clustering \citep{James2003}, K-centres functional clusterings \citep{Chiou2007}, functional principal analysis \citep{Dong2017}. We refer the reader to elsewhere \citep{Jacques2014b} for a more complete review of functional data clustering. On the other hand, clustering methods via Bayesian inference are developed. These methods include Bayesian latent class models or finite mixture models \citep{Neelon2011,Lu2019} and Bayesian non-parametric models via Dirichlet process mixture \citep{Zeldow2021}. \par

Many modern clinical or cohort studies capture multiple longitudinal characteristics of patients, such as anthropometrics, clinical symptoms, environmental exposures and genetic profiles, each of which may be collected at irregular and sparse time points. While it becomes increasingly challenging, modeling these data simultaneously reveals co-existing development patterns of different markers or responses. When it comes to clustering multiple longitudinal trajectories, methods can be roughly divided into two categories, namely methods for the same types of outcomes (e.g. continuous) and methods for mixed types of outcomes (i.e. both continuous and discrete). For the former, example include multivariate model-based clustering using mixed effect models for continuous markers  \citep{Marshall2006, Villarroel2009,Proust-Lima2015}, multivariate functional clustering for continuous \citep{Jacques2014} or discrete markers  \citep{Lim2020a},  Bayesian finite mixtures for continuous markers  \citep{Leiby2009, Fruehwirth-Schnatter2010, Lu2022} and hidden Markov models for continuous markers  \citep{Xia2019}. For the latter, examples include group-based multi-trajectory analysis \citep{Nagin2018}, which is an extension of the group-based single trajectory analysis \citep{Nagin1999}, and Bayesian mixture modeling for discrete and continuous longitudinal data (BMM) \citep{Komarek2013,Komarek2014}.  \par 

Recently, a Bayesian consensus clustering (BCC) approach was developed for clustering multiple continuous longitudinal markers \citep{Lu2022}. In the current study, we extended this approach to analyze mixed-type (continuous, discrete and categorical) longitudinal data. Compared to existing methods, several key features make the proposed model appealing: (a) it allows simultaneously clustering of continuous, discrete and categorical longitudinal data,  (b) it allows each longitudinal marker to be collected from different sources with measurements taken at distinct sets of time points, (c) it relaxes the assumption that all markers have the same clustering structure by estimating the marker-specific (local) clusterings and overall (global) clustering. These features enhance the utility and interpretability of the proposed model when it is applied to clinical and epidemiological data. \par

The rest of this paper is organized as follows: In Section 2, we describe the proposed model for clustering mixed-type multivariate longitudinal data. In Section 3, we describe the specification of prior and sketch posterior computation. In Section 4, we implement the proposed BCC to analyze the data from the CHILD Cohort Study. In Section 5, we perform a simulation study to evaluate the performance of the proposed model and compare it to an existing model under several scenarios. In Section 6, we discuss our findings.

\section{Proposed BCC Model for Mixed-Type Longitudinal Data}
Under the finite mixture model framework, a class of BCC model for performing integrative clustering based on multi-source data (multiple datasets collected from distinct sources) is proposed \citep{Lock2013}. This model assumes that there are separate clustering for each data source and these separate clusterings adhere loosely to an overall consensus clustering. Under this framework, \citet{Lu2022} developed a BCC model for clustering multiple continuous longitudinal markers. In the current study, we extended the BCC model to continuous, discrete and categorical longitudinal data, which are commonly seen in many clinical studies. 

Suppose there are $R$ mixed-type (continuous, discrete and categorical) longitudinal markers. Let $L_{i,r} = 1,...,K$ denote the source-specific (local) cluster label for subject $i$ and marker $r$ for $r=1,...,R$, and $C_i = 1,...,K$ denote the overall (global) cluster label for subject $i$. Let $\bm{y}_{i,r} = (y_{i1,r}, ..., y_{in_{i,r},r})^\top$ denote the $r^{th}$ longitudinal marker for the $i^{th}$ subject, $n_{i,r}$ is the number of measurements, where $i=1,...,N$ and $r=1,...,R$. Let $y_{ij,r}$ denote the $j^{th}$ observation of the $i^{th}$ subject for the $r^{th}$ marker, $i=1,...,N, j=1,...,n_{i,r}, r=1,...,R$. The BCC assumes that $\bm{L}_{r} = (L_{1,r},...,L_{N,r})$ are dependent on $\bm{C} = (C_1,...,C_N)$ through $P(L_{i,r}=k|C_i) = \vartheta(k,C_i, \alpha_r)$, where $\alpha_r$ adjusts the dependence function $\vartheta(\cdot)$. The joint distribution of the $R$ features for individual $i$ can be written as a multivariate finite mixture model,  
\begin{align}
f(\bm{y}_{i,1},...,\bm{y}_{i,R}) = \sum_{k_1,...,k_R} \Pi_{k_1,...,k_R} f(\bm{y}_{i,1},...,\bm{y}_{i,R}|L_{i,1}=k_1,...,L_{i,R}=k_R)
\end{align}
where $\Pi_{k_1,...,k_R}= P(L_{i,1} = k_1,...,L_{i,R} = k_R)$ and $k_1 \in \{1,...,K\}$,..., $k_R \in \{1,...,K\}$. The joint distribution of local clusterings can be written as
\begin{align}
P(L_{i,1},...,L_{i,R}) = \sum_kP(L_{i,1},...,L_{i,R}|C_i=k)P(C_i=k)
\end{align}

The BCC assumes the source-specific clusterings are dependent on the overall clustering, i.e. $P(L_{i,r}=k|C_i) = \vartheta(k, C_i,\alpha_r)$, where $\alpha_r$ adjusts the dependence function $\vartheta(\cdot)$. The conditional model can be specified as 
\begin{equation}
P(L_{i,r}=k|\bm{y}_{i,r},C_i,\bm{\gamma}_{k,r}, \bm{\beta}_{ik}) \propto\vartheta(k, C_i,\alpha_r)f_{k,r}(\bm{y}_{i,r}|\bm{\gamma}_{k,r}, \bm{\beta}_{ik,r})
\end{equation}
where $\bm{\gamma}_{k,r}$ and $\bm{\beta}_{ik,r}$ denote the fixed effects and random effects for individual $i$. A simple form for $\vartheta(k,C_i,\alpha_r)$ can be used  \citep{Lock2013}, that is  $\vartheta(k,C_i,\alpha_r) = \alpha_r$ if $C_i = L_{i,r}$ and $\vartheta(k,C_i,\alpha_r) = (1- \alpha_r)/(K-1)$ otherwise. Therefore, $\alpha_r$ can also be viewed as an adherence parameter representing the degree of agreement between $L_{i,r}$ and $C_i$. 

Moreover, we assume $ f_{k,r}(\bm{y}_{i,r}|\bm{\gamma}_{k,r}, \bm{\beta}_{ik,r})$ is a distribution from the exponential family with the dispersion parameter $\phi_{k,r}$ and the fully specified mean function is given by 
\begin{equation}
h^{-1}_{k,r}(E(\bm{y}_{i,r}|\bm{\beta}_{ik,r},\bm{\gamma}_{k,r})) = \bm{\eta}_{ik,r} = \bm{x}_{i,r}^\top\bm{\gamma}_{k,r} +  \bm{Z}_{i,r}^\top\bm{\beta}_{ik,r}
\end{equation}
where $h_{k,r}^{-1}$  is a canonical link function for the mean of the  marker $r$ in cluster $k$ . $\bm{\eta}_{ik,r} = (\eta_{i1k,r},...,\eta_{in_{i,r}k,r})^\top$ is the linear predictor for marker $r$. $\bm{x}_{i,r}$ is a $p_r\times 1$ vector of predictors and $\bm{\gamma}_{k,r}$ are the corresponding coefficients, $\bm{Z}_{i,r}$ is a $q_r\times 1$ vector of predictors, and $\bm{\beta}_{ik,r}|\cdot \sim \text{MVN}(\bm{0},\bm{\Sigma}_{k,r})$. Moreover, 
\begin{equation}
P(\bm{y}_{i,r}|\bm{\beta}_{ik,r},\bm{\gamma}_{k,r}, \phi_{k,r}) = \prod_{j=1}^{n_{i,r}} \text{exp}\bigg\{ \frac{y_{ij,r}\eta_{ijk,r} - g_{k,r}(\eta_{ijk,r})}{\phi_{k,r}} + w_{k,r}(y_{ij,r},\phi_{k,r}) \bigg\} 
\end{equation}
where $g(\eta)$ and $w(y,\phi)$ are appropriate distribution-specific functions. For example, for marker with Gaussian distribution, $g(\eta) = \eta^2/2$ and $w(y,\phi) = \frac{\log(2\pi\phi)}{2} - \frac{y^2}{2\phi}$. Furthermore, for  three markers with Gaussian, Poisson and Binomial distributions, respectively, we can specify the models for the means: 
\begin{align*}
  \text{(1): }  E(\bm{y}_{i,1}|\bm{\gamma}_{k,1}, \bm{\beta}_{ik,1}) & = \bm{\eta}_{ik,1}  =   \bm{x}_{i,1}^\top\bm{\gamma}_{k,1} +  \bm{Z}_{i,1}^\top\bm{\beta}_{ik,1}  \\
  \text{(2): }    \text{log}(E(\bm{y}_{i,2}|\bm{\gamma}_{k,2}, \bm{\beta}_{ik,2})) & = \bm{\eta}_{ik,2}  =  \bm{x}_{i,2}^\top\bm{\gamma}_{k,2} +  \bm{Z}_{i,2}^\top\bm{\beta}_{ik,2} \\
  \text{(3): }   \text{logit}(E(\bm{y}_{i,3}|\bm{\gamma}_{k,3},\bm{\beta}_{ik,3})) & =  \bm{\eta}_{ik,3} = \bm{x}_{i,3}^\top\bm{\gamma}_{k,3} +  \bm{Z}_{i,3}^\top\bm{\beta}_{ik,3}
\end{align*}
where $i=1,...,N$, $j=1,...,n_{i,r}$ and $r=1,2,3$. The model also involves dispersion parameters $\phi_{k,r} = \sigma^2_{k,r}$, for Gaussian distribution.  The corresponding dispersion parameters for logistic regression and Poisson regression are both 1, for $r=1,...,R$ and $k=1,...,K$. Given this proposed model, the complete-data likelihood can be written as

\begin{align}
\begin{split}
 \mathbb{L}^{Bayes}(\bm{\Theta}| \bm{y}_i,\bm{\beta}_i; \bm{x}_{i})   & = \sum_{k_1,...,k_R}   \bigg(\sum_k P(L_{i,1}=k_1,...,L_{i,R}=k_R|C_i=k)P(C_i=k)\bigg) \\ & \times \prod_{r=1}^R f(\bm{y}_{i,r}|\bm{\gamma}_{ik_r,r},\bm{\beta}_{ik_r,r}; \bm{x}_{i,r})  \\
& = \sum_{k_1,...,k_R}   \bigg(\sum_k \prod_{r=1}^R \vartheta(L_{i,r}=k_r,C_i=k,\alpha_r) \pi_k \bigg)  \prod_{r=1}^R f(\bm{y}_{i,r}|\bm{\gamma}_{ik_r,r},\bm{\beta}_{ik_r,r}; \bm{x}_{i,r}) 
\end{split}
\end{align}

where  $\bm{y}_i = (\bm{y}_{i,1},...,\bm{y}_{i,R})$,  $\bm{x}_i = (\bm{x}_{i,1},...,\bm{x}_{i,R})$, $\bm{\beta}_i = (\bm{\beta}_{i,1},..., \bm{\beta}_{i,R})$  and $\bm{\Theta}= (\bm{\alpha}, \bm{\pi},\bm{\gamma},\bm{\Sigma},\bm{\sigma})$.

%Moreover, we assume that $y_{ij,r}, r=1,...,R$ are conditionally independent given the class membership $L_{i,r}$, that is $f(y_{ij,1},...,y_{ij,R}|L_{i,1}=k_1,...,L_{i,R}=k_R) = \prod_{r=1}^R f(y_{ij,r}|L_{i,r}=k_r) $. Suppose $y_{ij,r}$ is generated from $f(y_{ij,r}|L_{i,r} = k_r) = f_{k,r}(y_{ij,r}|\bm{\theta}_{k,r})$.  

\section{Bayesian inference}
In this section, we describe the prior specification for the proposed model. For computational convenience, we consider independent conjugate priors for the model parameters.  

\subsection{Prior specification}
\begin{itemize}
	\item $\alpha_r \sim \text{TBeta}(\delta_{1,r}, \delta_{2,r}, 1/K)$, where $\text{TBeta}$ denotes a truncated Beta distribution ranged from $[1/K, 1]$.  Choosing $\delta_{1,r}=\delta_{2,r}=1$ results in the prior for $\alpha_r$ uniformly distributing between $1/K$ and 1. One can also assume that $\alpha=\alpha_1 = ...=\alpha_R$, which implies that each marker adheres equally well to the overall clustering $C$. In such case, the prior is $\alpha \sim \text{TBeta}(\delta_{1}, \delta_{2}, 1/K)$
	\item $\bm{\pi}\sim \text{Dirichlet}(\bm{\varphi}_0)$. Choosing $\bm{\varphi}_0 = (1,...,1)$ reflects no priori information favouring one cluster over the other.  
	\item Prior distributions for cluster-specific parameters $\bm{\gamma}_{k,r}, \Sigma_{k,r}, \phi_{k,r}$, for $k=1,...,K$ and $r=1,...,R$. 
	\begin{itemize}
	\item $\bm{\gamma}_{k,r} \sim \text{MVN}(\bm{0},\bm{V}_{0k,r})$, where $\bm{V}_{0k,r}$ is a $m_{k,r} \times m_{k,r}$ variance-covariance matrix, and $m_{k,r}$ is the dimension of $\bm{\gamma}_{k,r}$. Weakly informative prior can be obtained by setting $\bm{V}_{0k,r}$ to a diagonal matrix with large positive diagonal elements, e.g. 1000. 
	\item $\Sigma_{k,r}^{-1} \sim \text{Wishart}(\lambda_{0k,r}, (\lambda_{0k,r}\bm{\Lambda}_{0k,r})^{-1} )$, where the prior for the Wishart distribution is parametrized such that the mean is $\bm{\Lambda}_{0k,r}^{-1}$. In special case when $\Sigma_{k,r}$ is a diagonal matrix, i.e. $\Sigma_{k,r} = \text{diag}(\zeta^2_{k1,r},...,\zeta^2_{kq_r,r})$, the prior is an inverse gamma $\zeta^2_{kj,r} \sim \text{IG}(c_{0k,r},d_{0k,r}), j=1,...,q_r$,  where $c_{0k,r}$ and $d_{0k,r}$ are the parameters of an inverse gamma distribution.
	\item $\phi_{k,r}$ is not constant only when the marker is a Gaussian distribution. In such case, $\phi_{k,r} = \sigma^2_{k,r}$. The prior is 
	 $\sigma^2_{k,r} \sim \text{IG}(a_{0k,r},b_{0k,r}) $, where $a_{0k,r}$ and $b_{0k,r}$ are the parameters of an inverse gamma distribution. By definition, $\phi_{k,r} = 1$ for markers follow Poisson or Binomial distribution, for $k=1,...,K$ and $r=1,...,R$.
 	\end{itemize}
\end{itemize}

\subsection{Posterior computation}
For posterior computation, we develop a Gibbs sampling scheme coped with Metropolis-Hastings algorithm to update the model parameters.  In this subsection, we sketched the posterior updates for each parameter. The detail of the posterior computation is provided in Section A of the supplementary material.  

After initializing the parameters, the updates can be performed through the following steps.  At the $s$ step of the iteration, 
 \begin{itemize}
	\item Update local cluster membership $ L_{i,r}^{(s)}$ given $\{\bm{y}_{i,r}, \bm{\Theta}_r^{(s-1)}, \alpha_r^{(s-1)}, C_i^{(s-1)}, \bm{\beta}_r^{(s-1)} \}$, for $i=1,...,N$ and  $r=1,...,R$.
	\item Update $\alpha_r^{(s)}$ given $\{ \bm{C}^{(s-1)}, \bm{L}^{(s)} \}$, for $r=1,...,R$.
	\item Update $ C_i^{(s)}$ given $\{ \bm{L}^{(s)},  \bm{\alpha}^{(s)}, \bm{\pi}^{(s-1)}\}$, for $i=1,...,N$.
	\item Update $\bm{\pi}^{(s)}$ given $ \bm{C}^{(s)}$.
	\item Update cluster-specific parameters $\bm{\Theta}^{(s)} = (\bm{\gamma}^{(s)},\bm{\Sigma}^{(s)}, \bm{\sigma}^{2(s)})$ and  $\bm{\beta}^{(s)}$. In particular, $\bm{\gamma}^{(s)}$ and $\bm{\beta}^{(s)}$ are updated via Metropolis-Hastings algorithm.
\end{itemize}

Each MCMC iteration will produce a realization of cluster membership for $\bm{C}$, $\bm{L}_1$,..., $\bm{L}_R$. We use the mode over all the MCMC samples as the point estimates for both overall and marker-specific clusterings. 

Label switching is a common phenomenon in mixture models. This problem arises due to both the likelihood and posterior distributions are invariant to permutations of the parameters. To address this issue, in the real data application and simulation study, we applied a post-processing algorithm to reorder the labels based on Kullback-Leibler divergence \citep{Stephens2000}. Convergence of MCMC was diagnosed using visual inspection of trace plots as well as using the Geweke statistics \citep{Geweke1991}. Briefly, the Geweke statistics determines whether a Markov chain convergence or not based on a test for equality of the means of the first and last part of a Markov chain (by default the first 10\% and the last 50\%). If the samples are drawn from the stationary distribution of the chain, the two means are equal and Geweke's statistic has an asymptotically standard normal distribution. 

To determine the number of clusters, \citet{Lock2013} proposed an empirical separability criterion for BCC, which selects the value of $K$ that gives maximum adherence to an overall clustering.  For each $K$, the estimated adherence parameters $\alpha_r \in [\frac{1}{K},1]$, $r=1,...,R$ are mapped to the unit interval by the linear transformation  $\alpha^{*}_r = \frac{K\alpha_r - 1}{K-1}$
thus $\alpha^{*}_r \in [0,1]$. One then selects the value of $K$ that results in the highest mean adjusted adherence $\bar{\alpha}^{*} = \frac{1}{R}\sum_{m=1}^R \alpha^{*}_r$. \par
 
\subsection{Model assessment and goodness of fit} 
To assess the model goodness of fit, we considered the posterior predictive check. This approach is performed by comparing the observed data with data replicated from the posterior predictive distribution. If the model fits the data well, the replicated data, denoted as $\bm{y}^{rep}$ should have similar distribution as the observed data $\bm{y}$.   Consider the discrepancy measure $T = T(\bm{y},\bm{\Phi})$, where $\bm{\Phi}$ denotes all model parameters and $T$ can be sample quantiles or residual-based measures. Following \citet{Gelman1996}, a natural discrepancy measure is the $\chi^2$, which can be defined as
$\chi^2(\bm{y},\bm{\Phi}) = \sum^n_{i=1} \frac{ (\bm{y}_i - E(\bm{y}_i|\bm{\Phi}))^2}{Var(\bm{y}_i|\bm{\Phi})} $. In our proposed BCC model, we adapted the $\chi^2$ measure as
\begin{equation}
T^{obs}(\bm{y},\bm{\Phi})= \sum^K_{k=1} \sum_{i=1}^N \sum_{j=1}^{n_{i,r}}   \sum_{r=1}^R \frac{ z_{ik,r} ||\bm{y}_{i,r} - h^{-1}_{k,r}(E(\bm{y}_{i,r}|\bm{\beta}_{ik,r},\bm{\gamma}_{k,r}))||^2}{\sigma_{k,r}^2} 
\end{equation}
where $z_{ik,r} = 1$ if $L_{i,r}=k$, and 0 otherwise. This measure can be computed at each MCMC step by treating class indicator $z_{ik,r}$ and random effects $\bm{\beta}_{ik,r}$ as observed data. 

Bayesian predictive p-value is the probability that the discrepancy measure based on predictive sample, $ T^{rep}(\bm{y}^{rep},\bm{\Phi})$ is more extreme than the observed measure $T^{obs}(\bm{y},\bm{\Phi})$. This quantity can be estimated by computing the proportion of draws in which $T^{rep} > T^{obs}$. A p-value close to 0.5 indicates the model provides a good fit to the data, whereas a p-value close to 0 or 1 indicates a poor fit.\par

\section{Application to the CHILD Cohort Study}
It is widely acknowledged that asthma is a heterogeneous disease with many different clinical manifestations, and is difficult to diagnose in preschool children. To date, there is still no gold-standard case definition for preschool asthma despite most adults persistent asthma beginning in the preschool period, and therefore no solid basis for developing targeted early-life interventions. Acquiring the ability to objectively identify asthma phenotypes using data-driven methods could assist with identifying high-risk children and providing timely implementation of targeted treatment for this population.   Integration of related longitudinal asthma traits simultaneously will highlight co-existing patterns of these traits thereby improving our understanding of the etiology and developmental trajectory of asthma. These phenotypes will also facilitate predicting long-term outcomes and promoting the discovery of endotypes

\subsection{Description of data} 
The current study included 187 participants who had at least one observation for wheeze, one observation for cough and one observation for FEV/FVC (in z-score) during the first 5 years of age. These participants contributed to a total of 622 observations for the wheeze marker, 609 observations for the cough marker and 398 for the FEV/FVC marker.  The median (min, max) number of measurements per individual were 3 (1, 5), 3 (1, 5) and 2 (1, 5) for wheeze, cough and FEV/FVC, respectively.  Baseline demographics and clinical information were presented in Table E1. 

The primary goal of the current analysis was to discover subgroups (i.e. phenotypes) of children in the CHILD Cohort Study who shared similar longitudinal asthma traits, based on wheeze and cough, and $\text{FEV}/\text{FVC}$ (Figure 1). We analyzed the data using our proposed BCC model. \par
 
\subsection{Model specification} 
We used a binomial distribution for wheezing status (yes vs. no) and coughing status (yes vs. no), and a Gaussian distribution for the FEV/FVC z-score, respectively. In consideration of the clinical utility of the clusters and their interpretation, we fit the BCC model up to five classes.  On the basis of the observed trajectory of each marker and due to a small number of time points for each subject, we chose a linear form for each marker to approximate the trajectories, i.e. $\bm{x}_{i,r} = (1,\bm{t}_{i,r})^\top$ for $r=1, 2, 3$. To model the deviation of the individual trajectory from the mean trajectory within cluster, we used a random intercept model for all markers, i.e. $\bm{Z}_{i,r} = \bm{1}_{n_{i,r}}$, for $r=1,2,3$.\par

For the hyper-parameters of prior distributions defined in Section 3.1, vague prior distributions were used to reflect no prior information regarding the values of these parameters. Specifically, for the hyper-parameters of the adherence parameters $\bm{\alpha}_r$, we set $\delta_{1,r} = 1$ and $\delta_{2,r} = 1$ for $r=1,2,3$. For the global clustering proportion $\bm{\pi}$, we set $\bm{\varphi}_0 = (1,...,1)$. For the fixed effect coefficients $\bm{\gamma}_{k,r}$, we set  $V_{0k,r} = 25\bm{I}_3$. For the variance-covariance matrix $\Sigma_{k,r}$, we set $c_{0k,r} =0.001$ and $d_{0k,r} = 0.001$, for $k=1,...,K$ and $r=1,2,3$. For the residual variance of the Gaussian distribution, we set $a_{0k,r} = 0.001$ and $b_{0k,r} = 0.001$.

We ran the model with 30000 iterations, discarded the first 10000 samples, and kept every 20th sample. This resulted in 1000 samples for each model parameter. We applied a post-processing algorithm to reorder the labels based on Kullback-Leibler divergence \citep{Stephens2000} to address the label-switching problem. Two MCMC chains with different initial values were used to ensure the model parameters converge to the same posterior distributions. The computational process took about 30 minutes on a 12th Gen Intel(R) Core(TM) i9-12900 desktop computer. 

\subsection{Analysis results} 
The mean adjusted adherence $\bar{\alpha}^{*}$ suggested that a model with $K=3$ allowed the marker-specific clusterings contribute most information (on average) to the overall clustering (i.e. $\bar{\alpha}^{*}$ was the largest) compared to the models with a different $K$ (Figure E1). The posterior mean and 95\% credible interval (CI) for cluster-specific parameters of a three-class model were presented in Table 1. 

\begin{table}{}
  \caption{Posterior mean and 95\% credible interval of cluster-specific parameters}
     \scalebox{0.8}{ \begin{threeparttable} 
      {\begin{tabular}{rlccc}
    \toprule
                      &                   & \multicolumn{1}{l}{Cluster 1 $(k=1)$} & \multicolumn{1}{l}{Cluster 2 $(k=2)$} & \multicolumn{1}{l}{Cluster 3 $(k=3)$} \\
\cmidrule{3-5}    \multicolumn{1}{l}{Wheeze $(r=1)$} & $\gamma_{1k,r}$   & \multicolumn{1}{l}{1.74 (1.07, 2.70)} & \multicolumn{1}{l}{-1.95 (-2.45, -1.36)} & \multicolumn{1}{l}{-5.39 (-10.02, -2.74)} \\
                      & $\gamma_{2k,r}$   & \multicolumn{1}{l}{0.02 (-0.009, 0.07)} & \multicolumn{1}{l}{-0.04 (-0.06, -0.02)} & \multicolumn{1}{l}{-0.06 (-0.23, 0.13)} \\
                      & $\Sigma_{11k,r}^{*}$ & \multicolumn{1}{l}{1.12 (0.59, 2.23)} & \multicolumn{1}{l}{0.76 (0.42, 1.49)} & \multicolumn{1}{l}{1.11 (0.46, 2.51)} \\
                      &                   &                   &                   &  \\
    \multicolumn{1}{l}{Cough $(r=2)$} & $\gamma_{1k,r}$   & \multicolumn{1}{l}{0.02 (-0.33, 0.41)} & \multicolumn{1}{l}{-2.22 (-3.0, -1.43)} & \multicolumn{1}{l}{-6.12 (-11.0, -2.38)} \\
                      & $\gamma_{2k,r}$   & \multicolumn{1}{l}{-0.008 (-0.03, 0.02)} & \multicolumn{1}{l}{-0.05 (-0.07, -0.02)} & \multicolumn{1}{l}{0.009 (-0.15, 0.19)} \\
                      & $\Sigma_{11k,r}^{*}$ & \multicolumn{1}{l}{0.98 (0.53, 1.79)} & \multicolumn{1}{l}{0.73 (0.43, 1.26)} & \multicolumn{1}{l}{1.01 (0.46, 2.02)} \\
                      &                   &                   &                   &  \\
    \multicolumn{1}{l}{FEV/FVC z-score $(r=3)$} & $\gamma_{1k,r}$   & \multicolumn{1}{l}{-0.40 (-0.66, -0.16)} & \multicolumn{1}{l}{-0.007 (-0.17, 0.18)} & \multicolumn{1}{l}{0.55 (0.31, 0.79)} \\
                      & $\gamma_{2k,r}$   & \multicolumn{1}{l}{-0.01 (-0.03, -0.0007)} & \multicolumn{1}{l}{0.008 (0.001, 0.02)} & \multicolumn{1}{l}{-0.04 (-0.05, -0.03)} \\
                      & $\Sigma_{11k,r}^{*}$ & \multicolumn{1}{l}{1.02 (0.53. 1.92)} & \multicolumn{1}{l}{0.73 (0.43, 1.28)} & \multicolumn{1}{l}{1.03 (0.49, 2.07)} \\
                      & $\sigma^2_{k,r}$  & \multicolumn{1}{l}{0.69 (0.58, 0.82)}  & \multicolumn{1}{l}{0.69 (0.58, 0.82)}  & \multicolumn{1}{l}{0.69 (0.58, 0.82)} \\
    \bottomrule
    \end{tabular}}
    
     \begin{tablenotes} 
        \item  $\Sigma_{11k,r}^{*}$ denote the diagonal element of the variance-covariance matrix $\Sigma_{k,r}$ (in this model it only has one element) and is in the scale of $10^{-4}$, for $k=1,2,3$ and $r=1,2,3$. $\sigma^2_{k,r}$ was assumed to be identical across all clusters, i.e. $\sigma^2_{r} = \sigma^2_{1,r} = \sigma^2_{2,r} = \sigma^2_{3,r} $. 
   \end{tablenotes}	  
	\end{threeparttable}}
\end{table}
   
\begin{figure}[htbp]  
 \centering
   \includegraphics[width=12cm,height=8cm]{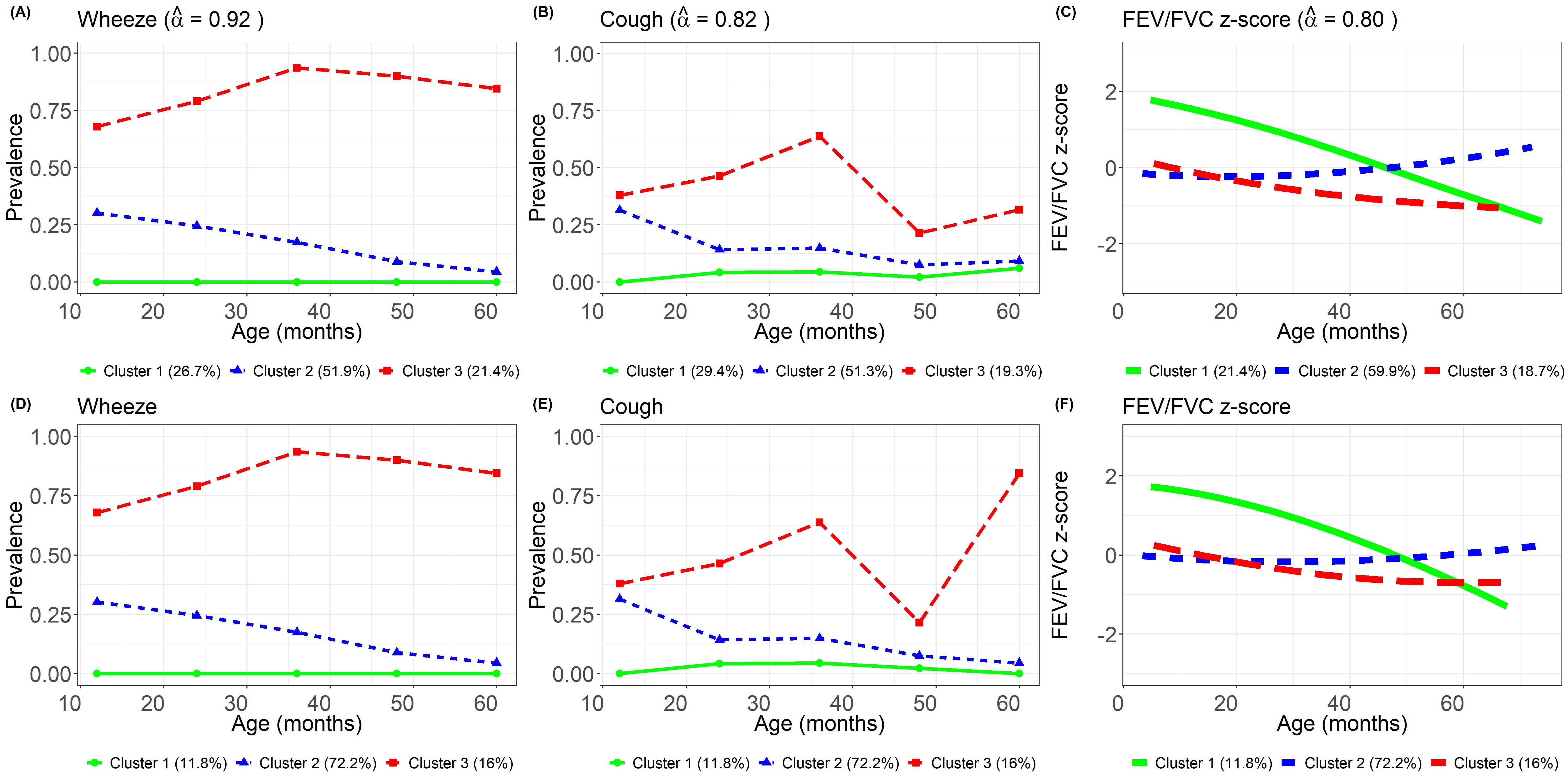}
\caption{Clustering results based on a three-class model. (A) Prevalence of wheezing over time based on wheeze-specific clustering, $\bm{L}_1$. (B) Prevalence of cough over time based on cough-specific clustering, $\bm{L}_2$. (C) FEV/FVC z-score trajectories based on FEV/FVC z-score-specific clustering, $\bm{L}_3$. (D) Prevalence of wheeze over time based on global clustering, $\bm{C}$. (E) Prevalence of cough over time based on global clustering, $\bm{C}$. (F) FEV/FVC z-score trajectories based on global clustering, $\bm{C}$.}
\end{figure}

\begin{figure}[htbp]  
 \centering
   \includegraphics[width=15cm,height=6cm]{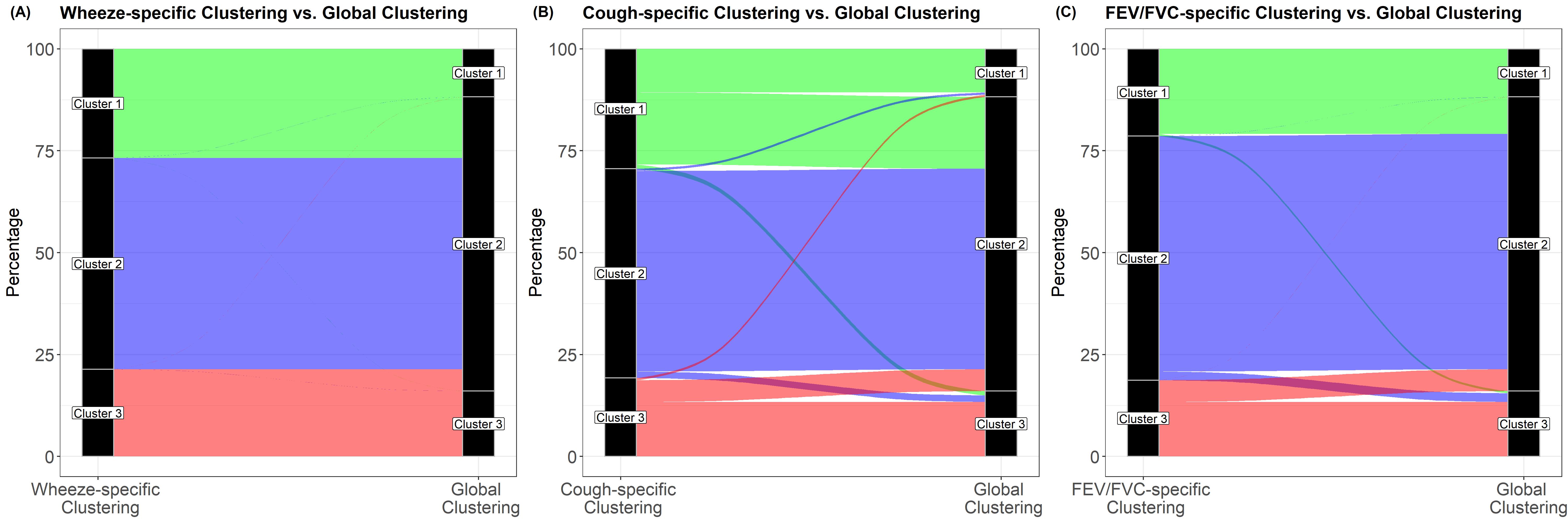}
\caption{Relationship between marker-specific clusterings and global clustering. (A) wheeze-specific clustering, $\bm{L}_1$ vs. global clustering, $\bm{C}$. (B) cough-specific clustering, $\bm{L}_2$  vs. global clustering, $\bm{C}$. (C) FEV/FVC-specific clustering, $\bm{L}_3$ vs. global clustering, $\bm{C}$.}
\end{figure}

Three distinct phenotypes based on wheeze, cough and lung function (FEV/FVC) trajectories were identified (Figure 2). These three marker-specific clusterings strongly adhered to the overall clustering with $\hat{\alpha}_1 = 0.92$ (95\%CI: 0.78, 1),   $\hat{\alpha}_2 = 0.82$ (95\%CI: 0.63, 0.98), $\hat{\alpha}_3 = 0.80$  (95\%CI: 0.50, 0.99),  respectively. Trace plots for the parameters $\bm{\alpha}$ are presented in Figure E2, indicating the two chains mixing well.  Specifically, local clustering of wheeze (i.e. wheeze-specific clustering) $\bm{L}_1$ (Figure 2A) revealed three distinct wheeze patterns over the first 60 months of age.  Cluster 1 (26.7\%) represented a group of children with infrequent or no wheezing. This group of children presented no wheezing symptoms over the first 60 months of age. Cluster 2 (51.9\%) represented a group of children with transient wheezing.  These children presented about 25\% of wheezing in the first 12 months with a decreasing trend and resolved completely by the time of 60 months.  Cluster 3 (21.4\%) represented a group of children with persistent wheezing. These children had a high prevalence of wheezing over the first 60 months of age. Similarly, local clustering of cough (i.e. cough-specific clustering) $\bm{L}_2$ (Figure 2B) revealed three distinct patterns of cough, namely Cluster 1 (29.4\%) representing infrequent/no coughing, Cluster 2 (51.3\%) representing transient coughing and Cluster 3 (19.3\%) representing persistent coughing. For the local clustering of FEV/FVC z-score (FEV/FVC-specific clustering) $\bm{L}_3$ (Figure 2C), Cluster 1 (21.4\%) represented a group of individuals with the best lung function at an early age (e.g. before 20 months) among the three clusters, but decreased over the course of the first 60 months. Cluster 2 (59.9\%) represented a group of individuals with stable lung function throughout the first 60 months, whereas Cluster 3 (18.7\%) represented a group of individuals with the worst lung function and with a decreasing trend.  When plotting by the global clustering $\bm{C}$, the trajectory patterns of wheeze, cough and FEV/FVC z-score remained similar to those by the local clusterings, but the cluster proportions were different (Figure 2 D, E, F). The proportions of Clusters 1, 2 and 3 according to global clustering $\bm{C}$ were 11.8\%, 72.2\% and 16\%, respectively. The relationship between marker-specific clusterings and global clustering was demonstrated in Figure 3, which indicated the change in the cluster membership from marker-specific clusterings to global clustering. This also suggests that global clustering integrates information from each marker to obtain an overall consensus clustering. \par 

To evaluate the clinical utility of the clusterings, we examined the association between the physician diagnosis of asthma at 5 years of age and the resulting marker-specific and global clusters, by computing the prevalence (i.e. sample proportion) of asthma within each cluster  (Figure 4). Specifically, when clustering based on wheezing symptoms only (wheeze-specific clustering), the prevalence of asthma in Clusters 1, 2 and 3 are 4.2\%, 7.3\% and 69.7\%, respectively (Figure 4A). When clustering based on coughing symptoms only (wheeze-specific clustering), the prevalence of asthma in Clusters 1, 2 and 3 are 7.5\%, 11.8\% and 54.8\%, respectively (Figure 4B). When clustering based on lung function only (FEVFVC-specific clustering), the prevalence of asthma in Clusters 1, 2 and 3 are 4.2\%, 7.3\% and 69.7\%, respectively (Figure 4C).  The overall clustering integrated information from wheeze, cough and FEV/FVC z-score (Figure 4D), which resulted in a more refined clustering. In particular, Cluster 1 represented a group of children with the lowest prevalence of asthma  (4.8\%). Cluster 2 represented a group of children with a moderate prevalence of asthma (8.5\%), and Cluster 3 represented a group of children with the highest prevalence of asthma (74.1\%), which was a more severe phenotype compared to Cluster 3 from the marker-specific clusterings. Moreover, these clusters were highly associated with gene and environmental risk factors (Figure E3). For example, percentages of individuals who visited the emergency room during the first 5 years were 0\%, 16.3\% and 91.7\% in Clusters 1, 2 and 3, respectively. This suggested Cluster 3 represented a group of individuals with the worst disease condition and demonstrated the utility of the resulting global clustering in aiding in making clinical decisions.  \par 

\begin{figure}[htbp]  
 \centering
   \includegraphics[width=15cm,height=8cm]{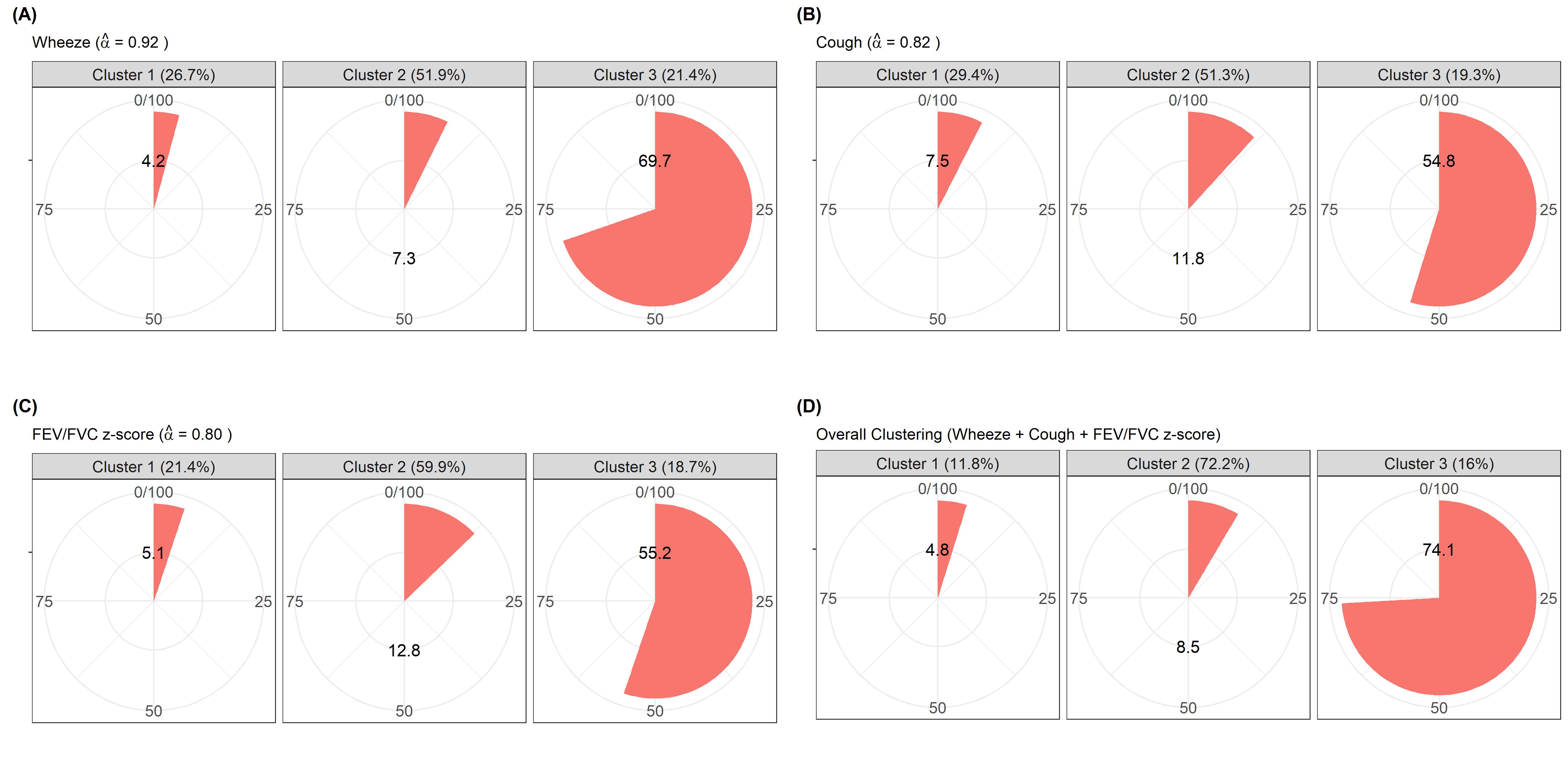}
\caption{Asthma prevalence at 5 years of age by local clusterings and global clusterings. (A) Asthma prevalence by wheeze-specific clustering $\bm{L}_1$.  The prevalence of asthma in Clusters 1, 2 and 3 are 4.2\%, 7.3\% and 69.7\%, respectively.  (B) Asthma prevalence by cough-specific clustering $\bm{L}_2$. The prevalence of asthma in Clusters 1, 2 and 3 are 7.5\%, 11.8\% and 54.8\%, respectively. 
 (C) Asthma prevalence by FEV/FVC z-score-specific clustering $\bm{L}_3$. The prevalence of asthma in Clusters 1, 2 and 3 are 5.1\%, 12.8\% and 55.2\%, respectively. (D) Asthma prevalence by global clustering $\bm{C}$. The prevalence of asthma in Clusters 1, 2 and 3 are 4.8\%, 8.5\% and 74.1\%, respectively. }
\end{figure}

Next, we investigated the goodness of fit and stability of the clusterings. To evaluate the model fitness, we computed the replicated T4 and observed T4 values for each child and calculated the Bayesian p values (0.3) (Figure E4). These results suggested that there was no clear evidence that the model provided a poor fit to the current data. To evaluate the stability of the clusterings, we used a cross-validation method. We focused on global clustering as it is the clustering of interest in practice. A stable clustering suggests individual cluster membership should not change if the data is changed in a non-essential way \citep{Hennig2007}. To this end, we created 50 subsets from the original data. Each dataset was generated by holding out 3 or 4 subjects. We refit the model and compared the overall clusterings to the one obtained from the model based on the original data. The agreement between the two clusterings (i.e. clustering based on the original dataset and the clustering based on a subset of the original data) was measured using the adjusted rand index (aRand) \citep{Hubert1985} and the Jaccard index \citep{Hennig2007}.  Higher aRand and Jaccard index suggest a better agreement between the two clusterings under evaluation.  The results were shown in Figure E5. All datasets yielded over 0.8 aRand and the lowest value was 0.83 (dataset 24) (Figure E5A). Similarly, all datasets yielded over 0.8 Jaccard index and the lowest value was 0.84 (dataset 49) (Figure E5B). Therefore, the aRand and Jaccard suggested a high agreement between the results based on the original data and those based on the subset. Posterior cluster probability suggested that Cluster 1 was highly stable and all children in this cluster had a posterior cluster probability larger than 0.5 (Figure E5(C)). 

Several sensitivity analyses were also performed to evaluate the consistency of the modeling results under different scenarios. Specifically, we also analyzed these markers separately based on a single-marker model, in which each marker entered the model at a time (i.e. separate clustering). These results were presented in Figure E6. The proportion of clusters for each marker produced by separated clustering differed from those based on BCC model (joint clustering). In addition, the trajectory patterns of these clusters (in particular the wheeze marker) also differed between separate clustering and joint clustering. The prevalence of asthma for the single-marker model was presented in Figure E7. The difference in asthma prevalence between clusters from the single-marker model was less striking compared to those from the join clustering from the proposed BCC model. In addition, we also fitted the proposed BCC model with random intercept and random slope for each marker, which resulted in similar clustering results (Table E2). Moreover, we  varied the hyper-parameters of the prior distributions for $\bm{\alpha}$ and $\bm{\pi} $, from non-informative (e.g. $\phi = 1, \delta_1 = 1, \delta_2 = 1$) to highly informative  (e.g. $\phi = 10, \delta_1 = 5, \delta_2 = 1$), the results suggested the class proportions were generally robust to different priors (Table E3). \par

\section{Simulation}
We conducted a simulation study to compare the proposed BCC model with the standard Bayesian mixture model for multivariate longitudinal data in terms of clustering performance.

\subsection{Simulation set-up}
To fully evaluate the performance of the proposed model, we generated data from three types of responses ($R=3$), namely the continuous, count and binary responses. Let the global cluster membership $\bm{C} = (C_1,...,C_N)$ define $K$ clusters. We considered $K=2, 3, 4 $ in the current simulation study. These settings are commonly seen in clinical studies. For $K = 2$, we let $C_i=1$ for $i=1,...,100$ and $C_i = 2$ for $i=101,...,200$, respectively. For $K=3$, we further let $C_i=3$ for $i=201,...,300$. For $K=4$, we added an additional $C_i=4$ for $i=301,...,400$. The local cluster labels $L_{i,r} \in (1,...,K)$ were generated with probability $P(L_{i,r} = C_i)=\alpha_r$ for $i=1,...,N$ and $r=1,...,R$. We let each individual have 4 measurements,  with the first visit time $t_{i1,r}$ being 0 and the remaining three visit times,  $t_{i1,r}$,  $t_{i2,r}$,  $t_{i3,r}$ being random variables with uniform distributions on intervals (0, 15), (15, 25) and (25, 30). The time variable generated in such a way can also be analyzed by standard methods such as BMM \citep{Komarek2013,Komarek2014}, and therefore comparison between our proposed model and BMM was made possible. The longitudinal markers $\bm{y}_{i,r}$ were then generated from the exponential family distributions according to the model in (2). The random effects (both intercept and slope) were generated from multivariate normal distribution, i.e. $\bm{\beta}_{ik,r}\sim \text{MVN}(\bm{0},\bm{\Sigma}_{k,r})$. To examine the impact of misspecification of random effect, we also generated the random effects under the multivariate t-distribution with 5 degrees of freedom \citep{Komarek2013}. The true $\bm{\alpha} $ were set to be (a) $ \bm{\alpha} = (1/K, 1/K, 1/K)$, which represented there was no relationship between the local clusterings and overall clustering, (b) $ \bm{\alpha} = (0.8, 0.8, 0.8)$, which represented there was a strong relationship between the local clusterings and overall clustering, (c) $ \bm{\alpha} = (1, 1, 1)$, which represented there was a perfect  relationship between the local clusterings and overall clustering and $ \bm{\alpha} = (1, 1/K, 0.8)$, which represented the case when one marker had a perfect relationship with the overall clustering, one marker had no relationship with the overall clustering (i.e noise variable) and one marker was perfectly related to the overall clustering.  Simulation scenarios for $K=2, 3, 4$ are shown in Figures E8, E9 and E10, respectively. For each setting, we generated 50 datasets. \par

We compared the proposed model with the BMM for mixed-type longitudinal data \citep{Komarek2013}, which was implemented via the R \textit{mixAK} package. For BMM, we used a linear model with a random intercept and random slope for each marker. We employed the aRand to measure the agreement between the estimated class membership and the true class membership. The aRand for overall clustering was denoted as aRand.G and the aRand for marker-specific clusterings were denoted as aRand.I. Since BMM does not inherently provide overall and marker-specific clusterings as BCC, the overall clustering for BMM was determined by joint clustering of all three markers together, whereas the marker-specific clusterings were determined by three univariate clustering models (i.e. each marker entered the model separately).  To calculate the aRand.I for BMM, we computed the mean aRand across the three univariate clustering models.   

\subsection{Simulation results}
The simulation results under various settings of $K$ and $\bm{\alpha}$ are presented in Table 2. We first discuss the settings with random effects generated from a normal distribution. When $K=2$ and $\bm{\alpha} = (0.5, 0.5, 0.5)$, we observed that neither BCC or BMM was able to identify the overall clustering, both yielding aRand.G of 0(0.01). In such a setting, accurately estimating the local clusterings provided little information to determine the overall clustering, because the local clusterings adhered poorly to the overall clustering. However, BCC yielded a higher aRand.I than BMM and suggested that its marker-specific clustering performance was better than that of the BMM. When  $\bm{\alpha} = (0.8, 0.8, 0.8)$, we observed that BCC yielded higher aRand.G and aRand.I compared to BMM, which suggested that BCC outperformed BMM in terms of both overall clustering and marker-specific clusterings. When $\bm{\alpha} = (1, 1, 1)$, BCC fully recovered the overall clustering and the marker-specific clusterings, yielding aRand.G of 1 (0) and aRand.I of 1(0), respectively. BMM also recovered the overall clustering with aRand.G of 1(0), but yielded a lower aRand.I (0.88 (0.08)) compared to BCC. When $\bm{\alpha} = (1, 0.5, 0.8)$, we observed that BCC's overall clustering performance was distorted by the inclusion of the noise variable that contributed no information to determine the overall clustering. Therefore, poorer performance was observed compared to BMM. However, BCC yielded a better marker-specific clustering performance compared to BMM. When $K=3$, we observed that BCC outperformed BMM in terms of both overall clustering and marker-specific clusterings in all settings except when $\bm{\alpha}=(0.34, 0.34, 0.34)$, whereas when $K=4$, we observed that BCC outperformed BMM in terms of both overall clustering and marker-specific clusterings in all settings. For settings with random effect from $t$ distribution,  BCC generally yielded slightly lower aRand.G and aRand.I compared to the same settings with random effects from a normal distribution. However, when comparing BCC to the BMM, similar results to the normal distribution settings were observed and the BCC generally outperformed the BMM. \par

Finally, we computed the root mean square error (RMSE) of key model parameters under different simulation settings described above. These results were presented in Tables E4 to E9 and they suggested that the proposed model yielded reasonably well parameter estimates in recovering the true parameter values under various settings. 

\begin{table}{}
  \caption{Simulation study to compare between BCC and BMM for clustering multivariate mixed-type longitudinal data under random effects from Normal and t distributions}
    \scalebox{0.8}{ \begin{threeparttable} 
 \begin{tabular}{cllllll}
\toprule
                  &                   &                   & \multicolumn{2}{c}{Normal distribution} & \multicolumn{2}{c}{t distribution} \\
\midrule
                  & $ \bm{\alpha} $          & Index             & BCC               & BMM               & BCC               & BMM \\
\midrule
\multirow{8}[2]{*}{$K = 2$} & \multirow{2}[1]{*}{(0.5, 0.5, 0.5)} & aRand.G           & \textbf{0(0.01)}  & 0(0.01)           & 0(0.01)           & 0(0) \\
                  &                   & aRand.I           & \textbf{0.95(0.02)} & 0.88(0.08)        & \textbf{0.94(0.02)} & 0.88(0.07) \\
                  & \multirow{2}[0]{*}{(0.8, 0.8, 0.8)} & aRand.G           & \textbf{0.57(0.1)} & 0.25(0.17)        & \textbf{0.58(0.07)} & 0.28(0.15) \\
                  &                   & aRand.I           & \textbf{0.95(0.02)} & 0.87(0.06)        & \textbf{0.94(0.02)} & 0.87(0.07) \\
                  & \multirow{2}[0]{*}{(1, 1, 1)} & aRand.G           & \textbf{1(0)}     & 1(0)              & \textbf{1(0)}     & 1(0) \\
                  &                   & aRand.I           & \textbf{1(0)}     & 0.88(0.08)        & \textbf{0.99(0.01)} & 0.87(0.08) \\
                  & \multirow{2}[1]{*}{(1, 0.5, 0.8)} & aRand.G           & 0.72(0.24)        & \textbf{0.8(0.4)} & 0.71(0.2)         & \textbf{0.8(0.38)} \\
                  &                   & aRand.I           & \textbf{0.96(0.01)} & 0.88(0.08)        & \textbf{0.94(0.02)} & 0.88(0.1) \\
\midrule
\multirow{8}[2]{*}{$K = 3$} & \multirow{2}[1]{*}{(0.34, 0.34, 0.34)} & aRand.G           & 0(0)              & 0(0)              & 0(0.01)           & \textbf{0(0)} \\
                  &                   & aRand.I           & 0.64(0.02)        & \textbf{0.7(0.05)} & \textbf{0.62(0.02)} & 0.61(0.03) \\
                  & \multirow{2}[0]{*}{(0.8, 0.8, 0.8)} & aRand.G           & \textbf{0.56(0.06)} & 0.35(0.10)       & \textbf{0.53(0.05)} & 0.37(0.1) \\
                  &                   & aRand.I           & \textbf{0.70(0.02)} & 0.70 (0.05)       & \textbf{0.67(0.02)} & 0.6(0.05) \\
                  & \multirow{2}[0]{*}{(1, 1, 1)} & aRand.G           & \textbf{1(0)}     & 0.64(0.24)        & \textbf{0.97(0.02)} & 0.79(0.23) \\
                  &                   & aRand.I           & \textbf{0.89(0.02)} & 0.71(0.07)        & \textbf{0.82(0.02)} & 0.64(0.05) \\
                  & \multirow{2}[1]{*}{(1, 0.34, 0.8)} & aRand.G           & \textbf{0.92(0.09)} & 0.52(0.27)        & \textbf{0.87(0.09)} & 0.63(0.34) \\
                  &                   & aRand.I           & \textbf{0.73(0.03)} & 0.69(0.06)        & \textbf{0.7(0.03)} & 0.6(0.04) \\
\midrule
\multirow{8}[2]{*}{$K = 4$} & \multirow{2}[1]{*}{(0.25, 0.25, 0.25)} & aRand.G           & \textbf{0(0)}     & 0(0)              & \textbf{0(0)}     & 0(0.01) \\
                  &                   & aRand.I           & \textbf{0.64(0.02)} & 0.62(0.06)        & \textbf{0.6(0.03)} & 0.59(0.04) \\
                  & \multirow{2}[0]{*}{(0.8, 0.8, 0.8)} & aRand.G           & \textbf{0.54(0.07)} & 0.54(0.07)        & \textbf{0.52(0.06)} & 0.37(0.1) \\
                  &                   & aRand.I           & \textbf{0.7(0.02)} & 0.7(0.02)         & \textbf{0.65(0.03)} & 0.57(0.05) \\
                  & \multirow{2}[0]{*}{(1, 1, 1)} & aRand.G           & \textbf{1(0.01)}  & 0.65(0.16)        & \textbf{1(0.01)}  & 0.72(0.2) \\
                  &                   & aRand.I           & \textbf{0.91(0.01)} & 0.61(0.08)        & \textbf{0.87(0.02)} & 0.58(0.04) \\
                  & \multirow{2}[1]{*}{(1, 0.25, 0.8)} & aRand.G           & \textbf{0.96(0.04)} & 0.23(0.32)        & \textbf{0.95(0.07)} & 0.49(0.33) \\
                  &                   & aRand.I           & \textbf{0.74(0.03)} & 0.62(0.05)        & \textbf{0.71(0.03)} & 0.59(0.03) \\
\bottomrule
\end{tabular}
   \begin{tablenotes} 
       \item Results are presented as mean (standard deviation) over 50 simulated datasets. aRAND.G: adjusted rand index between the true and estimated overall clusterings;  aRand.I: adjusted rand index between the true and estimated marker-specific clusterings; BCC: Bayesian consensus clustering; BMM: Bayesian mixture model; The aRand.G for BMM was calculated by multivariate clustering of all three markers together. The aRand.I for BMM was calculated by taking the average of aRand across the three single-marker models. The best results between BCC and BMM for each setting are in bold. 
   \end{tablenotes}	  
	\end{threeparttable}}
\end{table}

\section{Discussion}
In the current study, we developed a joint modeling approach within the BCC framework to analyze continuous, discrete and categorical longitudinal data. This type of data is commonly seen in clinical and epidemiological studies, and appropriate tools for analyzing such kind of data are in great need. The proposed model represents a novel model for clustering multivariate longitudinal data with mixed data types, with distinct sets of time points and data collection frequency. The flexibility and utility of the BCC model were demonstrated in the data from the CHILD Cohort Study and the simulation study. 

In analyzing the data from the CHILD Cohort Study, we discovered three distinct phenotypes in children characterized by distinct trajectory patterns of wheeze, cough and lung function (FEV/FVC z-score) over the first 60 months of age. The proposed BCC model provides an approach to integrating information from multiple longitudinal markers. Through integrative clustering using BCC, we found that children in Cluster 3 were associated with the highest risk of developing asthma compared to other clusters.   \par

Our simulation study demonstrated that the proposed BCC model yielded better clustering performance than BMM for mixed-type longitudinal markers when the data were generated from an underlying BCC model, particularly when the adherence between local and global clusterings is high. However, as it is seen in our simulation study, including a marker that has low adherence (i.e. $\alpha$ close to $1/K$) may distort the clustering performance. The current version of the model requires choosing the markers entering the model based on domain knowledge and clinical interest. However, automating the marker-selection process using data-driven approaches worth future studies.\par

Currently, several existing R packages can be used to implement cluster analysis for multiple longitudinal markers, such as K-mean clustering via \textit{kml3D} package \citep{Genolini2015}, which is an extension of  \textit{kml} package (for a single longitudinal marker), latent class mixed effect model via \textit{lcmm} package \citep{Proust-Lima2015} and Bayesian mixture model for multiple longitudinal markers via \textit{mixAK} package \citep{Komarek2014}. However, \textit{kml3D} package does not support discrete or categorical longitudinal markers. While \textit{lcmm} package allows clustering multiple continuous or multiple categorical (e.g. ordinal) markers, it does not support clustering mixed type markers (i.e. simoutenously clustering continuous and categorical markers), whereas  \textit{mixAK} package is one of the few packages can be used for clustering mixed-type longitudinal markers. To this end, our newly developed R package called \textit{BCClong} \citep{Tan2022} provides an alternative tool for clustering mixed-type longitudinal markers. This R package serves as a useful tool for implementing the proposed model described in the current study, particularly when the interests are in both marker-specific and global clusterings.  \par

The proposed model yields both maker-specific clusterings and global clustering. In practice, global clustering is often of greater interest and can be used to associate with health outcomes and exposures as well as predict long-term outcomes. Conceptually, marker-specific clustering can be viewed as the clustering based on a single marker alone, whereas global clustering encapsulates information from all markers and can be viewed as a weighted "average" clustering across markers, with weights being the adherence parameters. In order to capture the complexity of the disease, it is of great importance to consider multiple makers of interest simultaneously, as we often do for cross-sectional data. Reporting both marker-specific clusterings and global clustering may be useful in understanding the heterogeneity underlining each marker and the heterogeneity across markers. In addition, it also facilitates the understanding of the added values of the global clustering compared to marker-specific clusterings, particularly when they are associated with the outcome of interest. The adherence parameter $\bm{\alpha}$ reflects the contribution of each marker to the global clustering and therefore the overall heterogeneity. Practically,  the adherence parameter suggests to which extent a marker contributes to defining patients' asthma phenotypes and therefore serves as a measure of variable importance. 

Determining the number of clusters is a difficult step in model-based clustering and no widely accepted methods thus far. In general, we recommend determining the number of clusters based on both clinical and statistical considerations. From a clinical perspective, the chosen number of clusters should be clinically interpretable and meaningful. For example, they indicate disease severity and are associated with clinical outcomes and therefore can inform clinical decisions. From a statistical perspective, the chosen number of clusters should provide a good fit to the data, and result in stable clustering as well as parameter estimation.  Our model considered the number of classes $K$ as a fixed quantify and employed the mean adjusted adherence to determine the number of clusters. This approach yields a model in which all markers will on average contribute the most information to the overall clustering. However, the theoretical justification for this method is lacking and other methods for determining the number of clusters could be considered in the future study. Of note,  similar to the mean adjusted adherence criterion, other methods that are commonly used in practice also require a priori specification of the number of classes, such as the deviance information criterion, the Bayesian information criterion and the widely applicable information criterion. Alternatively, one could treat $K$ as a random quantity and infer it from the data. This may be achieved by using a Dirichlet process mixture \citep{Escobar1995} or a mixture of finite mixtures model \citep{Miller2018}. 

The proposed BCC model for multivariate longitudinal data borrows the idea from integrative clustering \citep{Richardson2016}, which refers to a class of cluster analysis methods that allow the integration of data from multiple sources with distinct structures. Such type of data is also known as multi-view data in the machine learning literature. Many of these methods were developed for cross-sectional omics (genomics, transcriptomics, proteomics) data to discover disease (e.g. cancer) subtypes but have not been adapted to longitudinal data settings. Our study represents a novel model and application of integrative clustering to a longitudinal data setting. While our study considers multiple longitudinal markers, future studies may consider extending this model to analyze multiple longitudinal datasets. To the best of our knowledge, such models have not been developed in the statistical literature. On the other hand, the underlying assumption of the proposed model is that there exists an overall clustering and that the local clusterings adhere to the overall clustering. However, statistical methods to examine such an assumption have not been developed. 

Several directions can be considered in future studies. For example, incorporating variable selection prior (e.g. spike and slab prior) \citep{Lu2021a, Lu2021b} for choosing variables that could influence the class membership will further enhance the flexibility of the model. The proposed BCC model also assumes that the number of clusters $K$ are the same between local clusterings and overall clusterings, which may not be realistic in some applications. Relaxing these assumptions would further enhance the flexibility of the proposed model. Another possible extension is to jointly model the longitudinal markers and health outcome, which will account for the uncertainty of the class membership derived from the mixture model. Nonetheless, the proposed BCC model provides new insight and serves as a flexible tool for clustering mixed-type multivariate longitudinal data. 

\section*{Software}
The R package \textit{BCClong} is available through the Comprehensive R Archive Network (CRAN).

\section*{Acknowledgement}
 We thank the CHILD Cohort Study (CHILD) participant families for their dedication and commitment to advancing health research. CHILD was initially funded by CIHR and AllerGen NCE. Visit CHILD at childcohort.ca. 

\section*{Data Availability Statement}
The CHILD Cohort Study data that support the findings of this paper are not openly available due to study participant consent restrictions. Information on accessing CHILD Cohort Study data and samples is available here: https://childstudy.ca/for-researchers/data-access/.

\section*{Funding}
ZT is supported by a Dean's Doctoral Award from Queen's University. PS holds a Tier 1 CRC Chair in Pediatric Asthma and Lung Health.  ZL is supported by a Discovery Grant funded by the Natural Sciences and Engineering Research Council of Canada. 

\section*{Supplement}
Computational details and additional analysis results are available in the supplementary material.

%\bibliography{ms}

\begin{thebibliography}{}

\bibitem[Chiou and Li, 2007]{Chiou2007}
Chiou, J.-M. and Li, P.-L. (2007).
\newblock Functional clustering and identifying substructures of longitudinal
  data.
\newblock {\em Journal of the Royal Statistical Society: Series B (Statistical
  Methodology)}, 69(4):679--699.

\bibitem[Ding et~al., 2021]{Ding2021}
Ding, M., Chavarro, J.~E., and Fitzmaurice, G.~M. (2021).
\newblock Development of a mixture model allowing for smoothing functions of
  longitudinal trajectories.
\newblock {\em Statistical Methods in Medical Research}, 30(2):549--562.

\bibitem[Dong et~al., 2017]{Dong2017}
Dong, J.~J., Wang, L., Gill, J., and Cao, J. (2017).
\newblock Functional principal component analysis of glomerular filtration rate
  curves after kidney transplant.
\newblock {\em Statistical methods in medical research}, page 0962280217712088.

\bibitem[Escobar and West, 1995]{Escobar1995}
Escobar, M.~D. and West, M. (1995).
\newblock Bayesian density estimation and inference using mixtures.
\newblock {\em Journal of the american statistical association},
  90(430):577--588.

\bibitem[Fr{\"u}hwirth-Schnatter and Pyne, 2010]{Fruehwirth-Schnatter2010}
Fr{\"u}hwirth-Schnatter, S. and Pyne, S. (2010).
\newblock Bayesian inference for finite mixtures of univariate and multivariate
  skew-normal and skew-t distributions.
\newblock {\em Biostatistics}, 11(2):317--336.

\bibitem[Gelman et~al., 1996]{Gelman1996}
Gelman, A., Meng, X.-L., and Stern, H. (1996).
\newblock Posterior predictive assessment of model fitness via realized
  discrepancies.
\newblock {\em Statistica sinica}, pages 733--760.

\bibitem[Genolini et~al., 2015]{Genolini2015}
Genolini, C., Alacoque, X., Sentenac, M., and Arnaud, C. (2015).
\newblock kml and kml3d: R packages to cluster longitudinal data.
\newblock {\em Journal of Statistical Software}, 65:1--34.

\bibitem[Genolini and Falissard, 2010]{Genolini2010}
Genolini, C. and Falissard, B. (2010).
\newblock Kml: k-means for longitudinal data.
\newblock {\em Computational Statistics}, 25(2):317--328.

\bibitem[Geweke, 1991]{Geweke1991}
Geweke, J. (1991).
\newblock {\em Evaluating the accuracy of sampling-based approaches to the
  calculation of posterior moments}, volume 196.
\newblock Federal Reserve Bank of Minneapolis, Research Department Minneapolis,
  MN, USA.

\bibitem[Hennig, 2007]{Hennig2007}
Hennig, C. (2007).
\newblock Cluster-wise assessment of cluster stability.
\newblock {\em Computational Statistics \& Data Analysis}, 52(1):258--271.

\bibitem[Hubert and Arabie, 1985]{Hubert1985}
Hubert, L. and Arabie, P. (1985).
\newblock Comparing partitions.
\newblock {\em Journal of classification}, 2(1):193--218.

\bibitem[Jacques and Preda, 2014a]{Jacques2014b}
Jacques, J. and Preda, C. (2014a).
\newblock Functional data clustering: a survey.
\newblock {\em Advances in Data Analysis and Classification}, 8(3):231--255.

\bibitem[Jacques and Preda, 2014b]{Jacques2014}
Jacques, J. and Preda, C. (2014b).
\newblock Model-based clustering for multivariate functional data.
\newblock {\em Computational Statistics \& Data Analysis}, 71:92--106.

\bibitem[James and Sugar, 2003]{James2003}
James, G.~M. and Sugar, C.~A. (2003).
\newblock Clustering for sparsely sampled functional data.
\newblock {\em Journal of the American Statistical Association},
  98(462):397--408.

\bibitem[Kom{\'a}rek et~al., 2013]{Komarek2013}
Kom{\'a}rek, A., Kom{\'a}rkov{\'a}, L., et~al. (2013).
\newblock Clustering for multivariate continuous and discrete longitudinal
  data.
\newblock {\em The Annals of Applied Statistics}, 7(1):177--200.

\bibitem[Kom{\'a}rek et~al., 2014]{Komarek2014}
Kom{\'a}rek, A., Kom{\'a}rkov{\'a}, L., et~al. (2014).
\newblock Capabilities of r package mixak for clustering based on multivariate
  continuous and discrete longitudinal data.
\newblock {\em Journal of Statistical Software}, 59(12):1--38.

\bibitem[Leiby et~al., 2009]{Leiby2009}
Leiby, B.~E., Sammel, M.~D., Ten~Have, T.~R., and Lynch, K.~G. (2009).
\newblock Identification of multivariate responders and non-responders by using
  bayesian growth curve latent class models.
\newblock {\em Journal of the Royal Statistical Society: Series C (Applied
  Statistics)}, 58(4):505--524.

\bibitem[Lim et~al., 2020]{Lim2020a}
Lim, Y., Cheung, Y.~K., and Oh, H.-S. (2020).
\newblock A generalization of functional clustering for discrete multivariate
  longitudinal data.
\newblock {\em Statistical methods in medical research}, 29(11):3205--3217.

\bibitem[Lock and Dunson, 2013]{Lock2013}
Lock, E.~F. and Dunson, D.~B. (2013).
\newblock Bayesian consensus clustering.
\newblock {\em Bioinformatics}, 29(20):2610--2616.

\bibitem[Lu and Lou, 2019]{Lu2019}
Lu, Z. and Lou, W. (2019).
\newblock Shape invariant mixture model for clustering non-linear longitudinal
  growth trajectories.
\newblock {\em Statistical methods in medical research}, 28(12):3769--3784.

\bibitem[Lu and Lou, 2021a]{Lu2021a}
Lu, Z. and Lou, W. (2021a).
\newblock Bayesian approaches to variable selection: a comparative study from
  practical perspectives.
\newblock {\em The International Journal of Biostatistics}.

\bibitem[Lu and Lou, 2021b]{Lu2021b}
Lu, Z. and Lou, W. (2021b).
\newblock Bayesian approaches to variable selection in mixture models with
  application to disease clustering.
\newblock {\em Journal of Applied Statistics}, pages 1--21.

\bibitem[Lu and Lou, 2022]{Lu2022}
Lu, Z. and Lou, W. (2022).
\newblock Bayesian consensus clustering for multivariate longitudinal data.
\newblock {\em Statistics in Medicine}, 41(1):108--127.

\bibitem[Marshall et~al., 2006]{Marshall2006}
Marshall, G., De~la Cruz-Mes{\'\i}a, R., Bar{\'o}n, A.~E., Rutledge, J.~H., and
  Zerbe, G.~O. (2006).
\newblock Non-linear random effects model for multivariate responses with
  missing data.
\newblock {\em Statistics in Medicine}, 25(16):2817--2830.

\bibitem[McNicholas and Subedi, 2012]{McNicholas2012}
McNicholas, P.~D. and Subedi, S. (2012).
\newblock Clustering gene expression time course data using mixtures of
  multivariate t-distributions.
\newblock {\em Journal of Statistical Planning and Inference},
  142(5):1114--1127.

\bibitem[Miller and Harrison, 2018]{Miller2018}
Miller, J.~W. and Harrison, M.~T. (2018).
\newblock Mixture models with a prior on the number of components.
\newblock {\em Journal of the American Statistical Association},
  113(521):340--356.

\bibitem[Nagin, 1999]{Nagin1999}
Nagin, D.~S. (1999).
\newblock Analyzing developmental trajectories: a semiparametric, group-based
  approach.
\newblock {\em Psychological methods}, 4(2):139.

\bibitem[Nagin et~al., 2018]{Nagin2018}
Nagin, D.~S., Jones, B.~L., Passos, V.~L., and Tremblay, R.~E. (2018).
\newblock Group-based multi-trajectory modeling.
\newblock {\em Statistical methods in medical research}, 27(7):2015--2023.

\bibitem[Neelon et~al., 2011]{Neelon2011}
Neelon, B., Swamy, G.~K., Burgette, L.~F., and Miranda, M.~L. (2011).
\newblock A bayesian growth mixture model to examine maternal hypertension and
  birth outcomes.
\newblock {\em Statistics in medicine}, 30(22):2721--2735.

\bibitem[Proust-Lima et~al., 2017]{Proust-Lima2015}
Proust-Lima, C., Philipps, V., and Liquet, B. (2017).
\newblock Estimation of extended mixed models using latent classes and latent
  processes: the r package lcmm.
\newblock {\em Journal of Statistical Software}, 78(2):1--56.

\bibitem[Richardson et~al., 2016]{Richardson2016}
Richardson, S., Tseng, G.~C., and Sun, W. (2016).
\newblock Statistical methods in integrative genomics.
\newblock {\em Annual review of statistics and its application}, 3:181--209.

\bibitem[Stephens, 2000]{Stephens2000}
Stephens, M. (2000).
\newblock Dealing with label switching in mixture models.
\newblock {\em Journal of the Royal Statistical Society: Series B (Statistical
  Methodology)}, 62(4):795--809.

\bibitem[Subbarao et~al., 2015]{Subbarao2015}
Subbarao, P., Anand, S.~S., Becker, A.~B., Befus, A.~D., Brauer, M., Brook,
  J.~R., Denburg, J.~A., HayGlass, K.~T., Kobor, M.~S., Kollmann, T.~R., et~al.
  (2015).
\newblock The canadian healthy infant longitudinal development (child) study:
  examining developmental origins of allergy and asthma.
\newblock {\em Thorax}, 70(10):998--1000.

\bibitem[Tan et~al., 2022]{Tan2022}
Tan, Z., Shen, C., and Lu, Z. (2022).
\newblock Bcclong package: Bayesian consensus clustering model for mixed-type
  longitudinal data.

\bibitem[Villarroel et~al., 2009]{Villarroel2009}
Villarroel, L., Marshall, G., and Bar{\'o}n, A.~E. (2009).
\newblock Cluster analysis using multivariate mixed effects models.
\newblock {\em Statistics in medicine}, 28(20):2552--2565.

\bibitem[Xia and Tang, 2019]{Xia2019}
Xia, Y.-M. and Tang, N.-S. (2019).
\newblock Bayesian analysis for mixture of latent variable hidden markov models
  with multivariate longitudinal data.
\newblock {\em Computational Statistics \& Data Analysis}, 132:190--211.

\bibitem[Zeldow et~al., 2021]{Zeldow2021}
Zeldow, B., Flory, J., Stephens-Shields, A., Raebel, M., and Roy, J.~A. (2021).
\newblock Functional clustering methods for longitudinal data with application
  to electronic health records.
\newblock {\em Statistical Methods in Medical Research}, 30(3):655--670.

\end{thebibliography}
%\bibliographystyle{apalike}  

\end{document}

\maketitle
\setcounter{figure}{0}
\renewcommand\thefigure{E\arabic{figure}}
\setcounter{table}{0}
\renewcommand\thetable{E\arabic{table}}

\section[A]{Details for Posterior Simulation}
In this section, we describe the details for updating the model parameters using Gibbs sampling scheme with Metropolis Hastings algorithm. 
\begin{itemize}
\item Generate local cluster membership $ L_{i,r}^{(s)}$ given $\{\bm{y}_{i,r}, \Theta_r^{(s)}, \alpha_r^{(s-1)}, C_i^{(s-1)} \}$, for $i=1,...,N$ and  $r=1,...,R$. The posterior probability that $L_{i,r}^{(s)} = k$ for $k=1,...,K$ is proportional to
	$\vartheta(L_{i,r}^{(s-1)} = k,C_i^{(s-1)}, \alpha_r^{(s-1)}) f_{k,r}(\bm{y}_{i,r}|\theta^{(s)}_{k,r}, \bm{\beta}_{i,r}) $
	\item Generate $\alpha_r^{(s)}$ given $\{ \bm{C}^{(s-1)}, \bm{L}^{(s)} \}$, for $r=1,...,R$. The posterior distribution for $\alpha_r^{(s)}$ is 
	$\text{TBeta}(\delta_{1,r} + \tau_r^{(s)}, \delta_{2,r}+N-\tau_r^{(s)},1/K)$
where $\tau_r^{(s)}$ is the number of samples satisfying $L^{(s)}_{i,r} = C^{(s-1)}_i$. If $\alpha=\alpha_1 = ...=\alpha_r$, the posterior distribution is 
	$\text{TBeta}(\delta_{1} + \tau^{(s)}, \delta_{2} + NR - \tau^{(s)}, 1/K)$
	where $ \tau^{(s)} = \sum_{r=1}^R\tau_r^{(s)}$.
	\item Generate $ C_i^{(s)}$ given $\{ \bm{L}^{(s)},  \bm{\alpha}^{(s)}, \bm{\pi}^{(s-1)}\}$. The posterior probability that $C_i^{(s)} = k$ is proportional to
$ \pi_k^{(s-1)}\prod_{r=1}^R \vartheta(C_i^{(s-1)} = k, L_{i,r}^{(s)}, \alpha_r^{(s)})$
	\item Generate $\bm{\pi}^{(s)}$ given $ \bm{C}^{(s)}$. The posterior distribution for $\bm{\pi}^{(s)}$ is $\text{Dirichlet}(\bm{\phi}_0+\bm{\rho})$, where $\rho_k$ is the number of samples allocated to cluster $k$ in $\bm{C}^{(s)}$.
	\item Generate cluster-specific parameters $\bm{\Theta}^{(s)} = (\bm{\gamma}^{(s)},\bm{\Sigma}^{(s)}, \bm{\sigma}^{2(s)})$ and  $\bm{\beta}^{(s)}$  given $\{{\bm{Y}, \bm{L}^{(s)}} \}$ \\
	 \begin{itemize}
	\item Update $\bm{\gamma}_{k,r}$ using Metropolis-Hastings algorithm. The full conditional distribution of the fixed effect parameter is 
	\begin{equation}
	f(\bm{\gamma}_{k,r}|\cdot) \propto \text{exp}\Big\{\phi_{k,r}^{-1}\sum_{i=1}^N z_{ik,r}(\bm{y}^{\top}_{i,r}\bm{\eta}_{ik,r} - \bm{1}^{\top}\bm{q}_{i,r}) - \frac{1}{2} \bm{\gamma}^{\top}_{k,r}\bm{V}^{-1}_{0k,r}\bm{\gamma}_{k,r} \Big\} 
	\end{equation}
where $\bm{\eta}_{ik,r} =  \bm{x}_{i,r}^\top\bm{\gamma}_{k,r} +  \bm{Z}_{i,r}^\top\bm{\beta}_{ik,r} $. Also,  $\bm{q}_{ik,r} =\bm{\eta}^2_{ik,r}/2$ for Gaussian,  $\bm{q}_{ik,r} = \log(1+\text{exp}(\bm{\eta}_{ik,r})) $  for Binomial and $\bm{q}_{ik,r} =\text{exp}(\bm{\eta}_{ik,r})$ for Poisson. In special case when the outcome is a Gaussian distribution, it can be shown that the above equation is a multivariate normal distribution with the variance and mean given by 
\begin{align}
\tilde{\bm{V}}_{k,r} = (\phi_{k,r}^{-1}\sum_{i=1}^N z_{ik,r}\bm{x}_{i,r}^{\top}\bm{x}_{i,r} + \bm{V}_{0k,r}^{-1})^{-1} 
\end{align}
\begin{align}
\tilde{\bm{v}}_{k,r} = \tilde{\bm{V}}_{k,r}\bigg(\phi_{k,r}^{-1}\sum_{i=1}^N z_{ik,r} \bm{x}_{i,r}^{\top} (\bm{y}_{i,r} - \bm{Z}_{i,r}^{\top}\bm{\beta}_{ik,r})\bigg) 
\end{align}

In general, when the outcome could be a non-Gaussian distribution, we use a quadratic approximation of $f(\bm{\gamma}_{k,r}|\cdot)$ to generate a candidate value of $\bm{\gamma}_{k,r}$. Assume that the maximum of $\log(f(\bm{\gamma}_{k,r}|\cdot))$ exists and that the Hessian is negative definite in the neighbourhood of $\bm{\hat{\gamma}}_{k,r}$ .  The gradient of partial derivatives of $\log(f(\bm{\gamma}_{k,r}|\cdot))$ is 
\begin{equation}
G(\bm{\gamma}_{k,r})= \frac{\partial \log(L^{Bayes})}{\partial \bm{\gamma}_{k,r}} = \phi^{-1}_{k,r}\sum_{i=1}^Nz_{ik,r}\bm{x}_{i,r}^{\top}(\bm{y}_{i,r} - \bm{\eta}_{ik,r})
\end{equation}
The negative of Hessian or matrix of the second derivative is 
\begin{equation}
H(\bm{\gamma}_{k,r})= -\frac{\partial^2 \log(L^{Bayes})}{\partial \bm{\gamma}_{k,r} \partial \bm{\gamma}_{k,r}^{\top}} =  \phi^{-2}_{k,r}\sum_{i=1}^Nz_{ik,r}\bm{x}_{i,r}^{\top}\kappa_{ik,r}\bm{x}_{i,r}
\end{equation}
where $\kappa_{ik,r}$ is a $n_{i,r} \times n_{i,r}$ dimensional diagonal matrix of the variance of the $r^{th}$ marker within the class $k$. For example, if the outcome has a Gaussian distribution, then the  $\kappa_{ik,r} =\sigma^2_{k,r} I_{n_{i,r} \times n_{i,r}}$, where $I_{n_{i,r} \times n_{i,r}}$ denotes an identity matrix with $n_{i,r} \times n_{i,r}$ dimension. The sampling steps are described as follow.
	\begin{enumerate}	
	\item A new value of $\bm{\gamma}^{*}_{k,r}$ is proposed by sampling from $\text{MVN}(\tilde{\bm{v}}_{k,r}^{(s)}, \tau_{k,r}\tilde{\bm{V}}_{k,r}^{(s)})$, where
		\begin{equation}
		\tilde{\bm{V}}_{k,r}^{(s)} =  \bigg(\bm{V}^{-1}_{0k} + H(\bm{\gamma}_{k,r}^{(s-1)}) \bigg)^{-1}  = \bigg(\bm{V}^{-1}_{0k} + \phi^{-2(s-1)}_{k,r}  \sum_{i=1}^N z_{ik,r}^{(s)}\bm{x}_{i,r}^\top \kappa_{ik,r}^{(s-1)}  \bm{x}_{i,r}   \bigg)^{-1}$$
		 $$\tilde{\bm{v}}_{k,r}^{(s)} = \bm{\gamma}_{k,r}^{(s-1)} +  \Delta_{\bm{\gamma},ik}^{(s)}
		\end{equation}	 
	where 
	\begin{align}
	\begin{split}
	\Delta_{\bm{\gamma},ik}^{(s)} &= \tilde{\bm{V}}_{k,r}^{(s)}   \bigg( G(\bm{\gamma}_{k,r}^{(s-1)}) - \bm{V}_{0k,r}^{-1} \bm{\gamma}_{k,r}^{(s-1)} \bigg) \\  &=  \tilde{\bm{V}}_{k,r}^{(s)}   \bigg(\phi^{-1(s-1)}_{k,r}\sum_{i=1}^N z_{ik,r}^{(s)} \bm{x}_{i,r}^\top (\bm{y}_{i,r} - \bm{\eta}_{ik,r}^{(s-1)}) - \bm{V}_{0k,r}^{-1} \bm{\gamma}_{k,r}^{(s-1)} \bigg) 
	\end{split}
	\end{align}	
 $\tau_{k,r} > 0$ is a tuning parameter. $\kappa_{ik,r} $ denotes the variance of marker $r$ for subject $i$. For Gaussian distribution $\kappa_{ik,r}  = \sigma^2_{ik,r}$, for Poisson distribution, $\kappa_{ik,r} = \text{exp}(\bm{\eta}_{ik,r})$, where $\lambda_{k,r}$ is the parameter of the Poisson distribution and for Binomial distribution, $\kappa_{ik,r}  = p_{ik,r}(1- p_{ik,r})$, where $p_{ik,r} = \text{exp}(\bm{\eta}_{ik,r})/(1+\text{exp}(\bm{\eta}_{ik,r}))$ is the parameter of the Binomial distribution.  

	\item Calculate the acceptance probability (the probability of move)
	 \begin{equation}
	 \alpha(\bm{\gamma}_{k,r}^{\text{new}},\bm{\gamma}_{k,r}^{(s)}) = \text{min} \Bigg\{1,\frac{f(\bm{\gamma}_{k,r}^{\text{new}}|\cdot)}{f(\bm{\gamma}_{k,r}^{(s)}|\cdot)} \frac{g(\bm{\gamma}_{k,r}^{\text{new}}|\bm{\gamma}_{k,r}^{(s-1)})}{g(\bm{\gamma}_{k,r}^{(s-1)}|\bm{\gamma}_{k,r}^{\text{new}})}\Bigg\} 
	\end{equation}	     
	\end{enumerate}	

	\begin{itemize}
	\item generate  a random variable $ u \sim \text{uniform}(0,1)$
	\begin{itemize}
	\item if $ u <  \alpha(\bm{\gamma}_{k,r}^{\text{new}},\bm{\gamma}_{k,r}^{(s)}) $ then accept the proposed values, i.e. set 
	$ \bm{\gamma}_{k,r}^{(s + 1)} = \bm{\gamma}_{k,r}^{\text{new}}$ 
	\item else reject the proposed values, i.e. set $ \bm{\gamma}_{k,r}^{(s + 1)} = \bm{\gamma}_{k,r}^{(s)}$ 
	\end{itemize}
	\end{itemize}
	Since $g(\cdot)$ is a symmetric distribution,  $\frac{g(\bm{\gamma}_{k,r}^{\text{new}}|\bm{\gamma}_{k,r}^{(s-1)})}{g(\bm{\gamma}_{k,r}^{(s-1)}|\bm{\gamma}_{k,r}^{\text{new}})} = 1$.

	\item Update $\Sigma_{k,r}^{-1*(s)}|\cdot \sim \text{Wishart}(\sum_{i=1}^K z_{ik,r}^{(s)} + \lambda_{0k,r}, \tilde{\bm{\Lambda}}_{k,r}^{(s)})$, \\ where $\tilde{\bm{\Lambda}}_{k,r}^{(s)}= (\lambda_{0k,r}\bm{\Lambda}_{0k,r}^{-1} + \sum_{i=1}^N z_{ik,r}^{(s)} \bm{\beta}_{ik,r}^{*(s-1)\top} \bm{\beta}_{ik,r}^{*(s-1)})^{-1}$. 

	\item Update random effect  $\bm{\beta}_{ik,r}$ using Metropolis-Hastings algorithm. 
	 The full conditional distribution of the random effect is 

	\begin{equation}
	f(\bm{\beta}_{ik,r}|\cdot) \propto \text{exp}\Big\{\phi_{k,r}^{-1}\sum_{i=1}^N z_{ik,r}(\bm{y}^{\top}_{i,r}\bm{\eta}_{ik,r} - \bm{1}^{\top}\bm{q}_{i,r}) - \frac{1}{2} \bm{\beta}^{\top}_{ik,r}\Sigma^{-1}_{k,r}\bm{\beta}_{ik,r} \Big\}  \\
	\end{equation}

In the special case when the outcome is a Gaussian distribution, it can be shown that the above equation is a multivariate normal distribution with the variance and mean given by 	
\begin{equation}
\tilde{\Sigma}_{k,r}^{(s)} = ({\Sigma^{-1(s)}_{k,r} + \frac{1}{\sigma_{k,r}^{2(s-1)}}\sum_{i=1}^N z_{ik,r}^{(s)} \bm{Z}_{i,r}^\top \bm{Z}_{i,r}})^{-1}
\end{equation}
\begin{equation}
\tilde{\bm{\mu}}_{k,r}^{(s)} = \tilde{\Sigma}_{k,r}^{(s)}\bigg(\frac{1}{\sigma^{2(s-1)}_{k,r}} \sum_{i=1}^N z_{ik,r}^{(s)} \bm{Z}_{i,r}^{\top}(\bm{y}_{i,r} - \bm{x}_{i,r}^{\top}\bm{\gamma}_{k,r}^{(s)})\bigg)
\end{equation}
	
In general, when the outcome could be a non-Gaussian distribution, we use a quadratic approximation of $f(\bm{\beta}_{ik,r}|\cdot)$ to generate a candidate value of $\bm{\beta}_{ik,r}$. Assume that the maximum of $\log(f(\bm{\beta}_{ik,r}|\cdot))$ exists and that the Hessian is negative definite in the neighbourhood of $\bm{\hat{\beta}}_{ik,r}$. The gradient of partial derivatives of $\log(f(\bm{\beta}_{ik,r}|\cdot))$ is 
\begin{equation}
G(\bm{\beta}_{ik,r})= \frac{\partial \log(L^{Bayes})}{\partial \bm{\beta}_{ik,r}} = \phi^{-1}_{k,r}\sum_{i=1}^Nz_{ik,r}\bm{Z}_{i,r}^{\top}(\bm{y}_{i,r} - \bm{\eta}_{ik,r}^{\top})
\end{equation}
The negative of Hessian or matrix of the second derivative is 
\begin{equation}
H(\bm{\beta}_{ik,r})= -\frac{\partial^2 \log(L^{Bayes})}{\partial \bm{\beta}_{k,r} \partial \bm{\beta}_{ik,r}^{\top}} =  \phi^{-2}_{k,r}\sum_{i=1}^Nz_{ik,r}\bm{Z}_{i,r}^{\top}\kappa_{ik,r}\bm{Z}_{i,r}
\end{equation}	
The sampling steps are described as follows. 
	\begin{enumerate}
	 \item A new value  $\bm{\beta}_{ik,r}^{\text{new}}$ is proposed, $\bm{\beta}_{ik,r}^{\text{new}}|\cdot \sim \text{MVN}(\tilde{\bm{\mu}}_{k,r}^{(s)}, \tau_{k,r}\tilde{\Sigma}_{k,r}^{(s)})$, where 
	\begin{equation}
	\tilde{\Sigma}_{k,r}^{(s)} =  (\Sigma^{-1(s)}_{k,r} + H(\bm{\beta}_{ik,r}^{(s-1)}))^{-1} = ({\Sigma^{-1(s)}_{k,r} + \phi^{-2 (s-1)}_{k,r}  \sum_{i=1}^N z_{ik,r}^{(s)} \bm{Z}_{i,r}^\top \kappa_{ik,r}^{(s-1)} \bm{Z}_{i,r}})^{-1}
	\end{equation}
	\begin{equation}
	\tilde{\bm{\mu}}_{k,r}^{(s)} = \bm{\beta}_{ik,r}^{(s-1)} + \Delta_{\bm{\beta},ik,r}^{(s)}  
	\end{equation}
	where 
	\begin{align}
	\begin{split}
	\Delta_{\bm{\beta},ik,r}^{(s)} &=  \tilde{\Sigma}_{k,r}^{(s)} \bigg( G(\bm{\beta}_{ik,r}^{(s-1)})  -  \Sigma_{k,r}^{-1}\bm{\beta}_{ik,r}^{(s-1)}\bigg)\\
	&= \tilde{\Sigma}_{k,r}^{(s)} \bigg( \phi^{-1 (s-1)}_{k,r}\sum_{i=1}^N z_{ik,r}^{(s)} \bm{Z}_{i,r}^{\top}(\bm{y}_{i,r} - \bm{\eta}_{ik,r}^{(s-1)}) -  \Sigma_{k,r}^{-1}\bm{\beta}_{ik,r}^{(s-1)}\bigg)
	\end{split}
	\end{align}
	where $ \tau_{k,r} > 0$ is a tuning parameter. We choose the value of $ \tau_{k,r}$ such that the acceptance rate is between 0.2 to 0.5.
	\item Calculate the acceptance probability 
	\begin{equation}
\alpha(\bm{\beta}_{ik,r}^{\text{new}},\bm{\beta}_{ik,r}^{(s)}) = \text{min} \Bigg\{1,\frac{f(\bm{\beta}_{ik,r}^{\text{new}}|\cdot)}{f(\bm{\beta}_{ik,r}^{(s)}|\cdot)} \frac{g(\bm{\beta}_{ik,r}^{\text{new}}|\bm{\beta}_{ik,r}^{(s)})}{g(\bm{\beta}_{ik,r}^{(s)}|\bm{\beta}_{ik,r}^{\text{new}})}\Bigg\} 
	\end{equation}
	 \end{enumerate}
where $f(\bm{\beta}_{ik,r}|\cdot)  \propto 
	\text{exp}\Big\{\phi_{k,r}^{-1}\sum_{i=1}^N z_{ik,r}^{(s)}(\bm{y}^{\top}_{i,r}\bm{\eta}_{ik,r} - \bm{1}^{\top}\bm{q}_{ik,r}) - \frac{1}{2} \bm{\beta}_{ik,r}^{\top}\Sigma^{-1}_{k,r}\bm{\beta}_{ik,r}\Big\}$. $\bm{\eta}_{ik,r} =  \bm{x}_{i,r}^\top\bm{\gamma}_{k,r} +  \bm{Z}_{i,r}^\top\bm{\beta}_{ik,r} $. Also,  $\bm{q}_{ik,r} =\bm{\eta}^2_{ik,r}/2$ for Gaussian,  $\bm{q}_{ik,r} = \log(1+\text{exp}(\bm{\eta}_{ik,r})) $  for Binomial and $\bm{q}_{ik,r} =\text{exp}(\bm{\eta}_{ik,r})$ for Poisson. Since the $g(\cdot)$ is symmetric, $\frac{g(\bm{\beta}_{ik,r}^{\text{new}}|\bm{\beta}_{ik,r}^{(s)})}{g(\bm{\beta}_{ik,r}^{(s)}|\bm{\beta}_{ik,r}^{\text{new}})} = 1$. 

	 \begin{itemize}
	\item generate  a random variable $ u \sim \text{uniform}(0,1)$
	\begin{enumerate}
	\item if $ u <  \alpha(\bm{\beta}_{ik,r}^{\text{new}},\bm{\beta}_{ik,r}^{(s)}) $ then accept the proposed values, i.e. set 
	$ \bm{\beta}_{ik,r}^{(s + 1)} = \bm{\beta}_{ik,r}^{\text{new}}$ 
	\item else reject the proposed values, i.e. set $ \bm{\beta}_{ik,r}^{(s + 1)} = \bm{\beta}_{ik,r}^{(s)}$ 
	\end{enumerate}
	\end{itemize}
	
	\item Update dispersion parameters $\phi_{k,r}^{(s)}$ for markers that are normally distributed, in such case, $\phi_{k,r}^{(s)} = \sigma^{2(s)}_{k,r}$ and we update 
	 $\sigma^{2(s)}_{k,r}|\cdot \sim \text{IG}(\tilde{a}_{0k,r}^{(s)},\tilde{b}_{0k,r}^{(s)})$, where $\tilde{a}_{0k,r}^{(s)} = a_{0k,r} + \frac{1}{2}\sum_{i=1}^N n_{i,r} z_{ik,r}^{(s)}$ and $\tilde{b}_{0k,r}^{(s)} = b_{0k,r} + \frac{1}{2}\sum_{i=1}^N z_{ik,r}^{(s)}||\bm{y}_{i,r} - \bm{x}_{i,r}^{\top}\bm{\gamma}_{k,r}^{(s)}  - \bm{Z}^{\top}_{i,r}\bm{\beta}_{ik,r}^{(s)}||^2$. 
	 When $\sigma^{2(s)}_{k,r}$ is assumed to be common across all cluster i.e.  $\sigma^{2(s)}_{1,r} = ... = \sigma^{2(s)}_{K,r} = \sigma^{2(s)}_{r}$, we update $\sigma^{2(s)}_{r}|\cdot \sim \text{IG}(\tilde{a}_{0,r}^{(s)},\tilde{b}_{0,r}^{(s)})$, where  $\tilde{a}_{0,r}^{(s)} = \sum_{k=1}^K \tilde{a}_{0k,r}^{(s)}$ and $\tilde{b}_{0,r}^{(s)} =  \sum_{k=1}^K \tilde{b}_{0k,r}^{(s)}$. For markers with Poisson or Binomial distribution,  $\phi_{k,r}=1$ for $k=1,...,K$ and $r=1,...,R$ by definition. 
	\end{itemize}
	\end{itemize}
	
\newpage
\section[B]{Supplementary Figures and Tables}
\begin{figure}[htbp]  
 \centering
   \includegraphics[width=11cm,height=11cm]{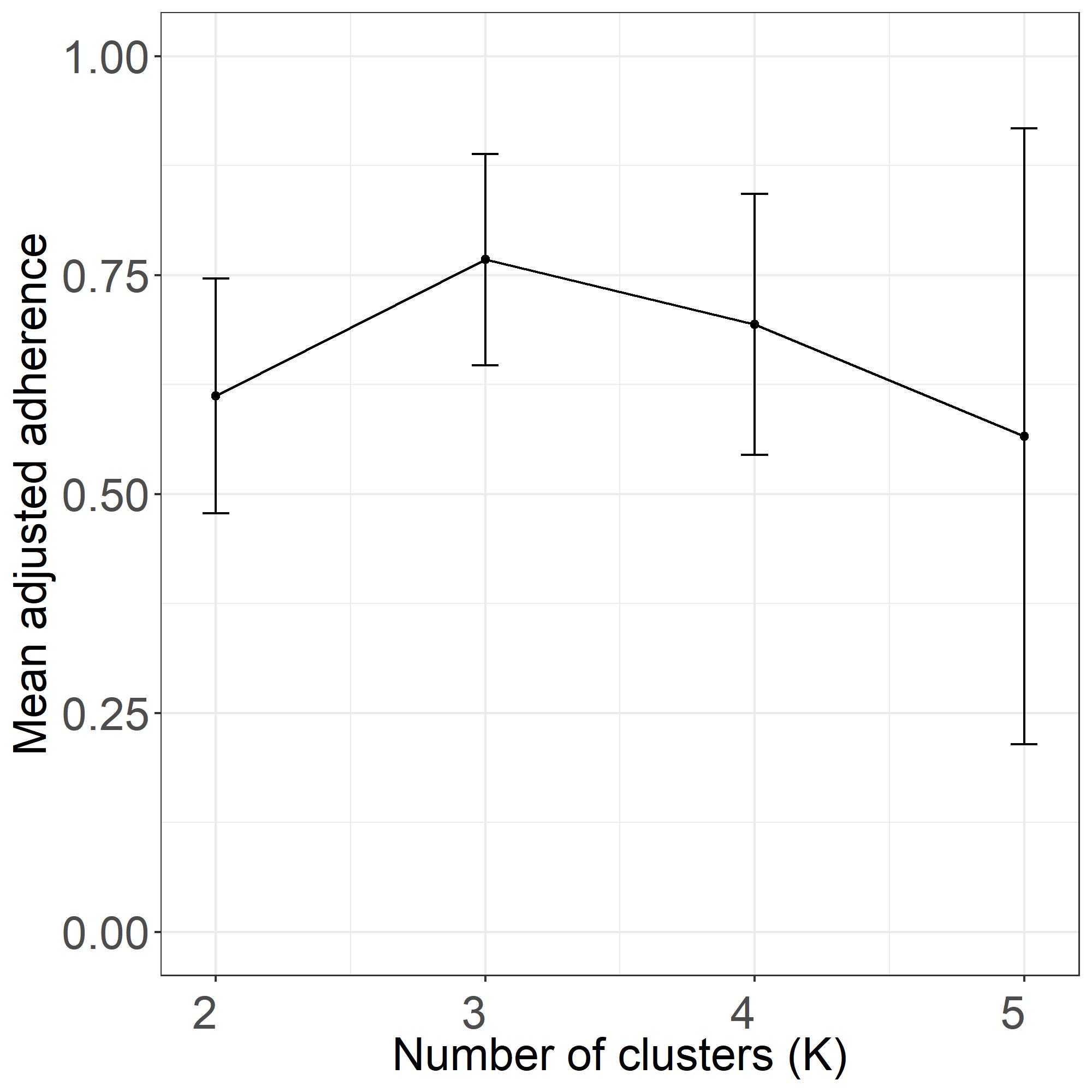}
\caption{Mean adjusted adherence for the different number of clusters.}
\end{figure}

\newpage
 \begin{figure}[htbp]  
 \centering
   \includegraphics[width=16cm,height=6cm]{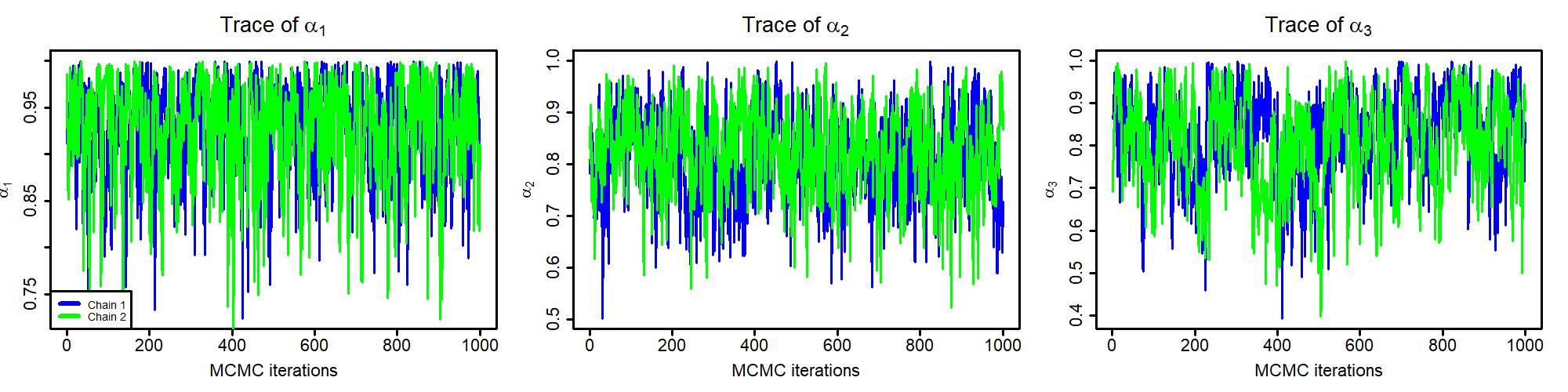}
\caption{Trace plots for $\alpha_1$, $\alpha_2$ and $\alpha_3$ based on two chains.}
\end{figure}

%\newpage
% \begin{figure}[htbp]  
% \centering
%   \includegraphics[width=16cm,height=6cm]{trace_plot_pi.JPEG}
%\caption{Trace plots for $\pi_1$, $\pi_2$ and $\pi_3$ based on two chains.}
%\end{figure}

\newpage
 \begin{figure}[htbp]  
 \centering
   \includegraphics[width=15cm,height=6cm]{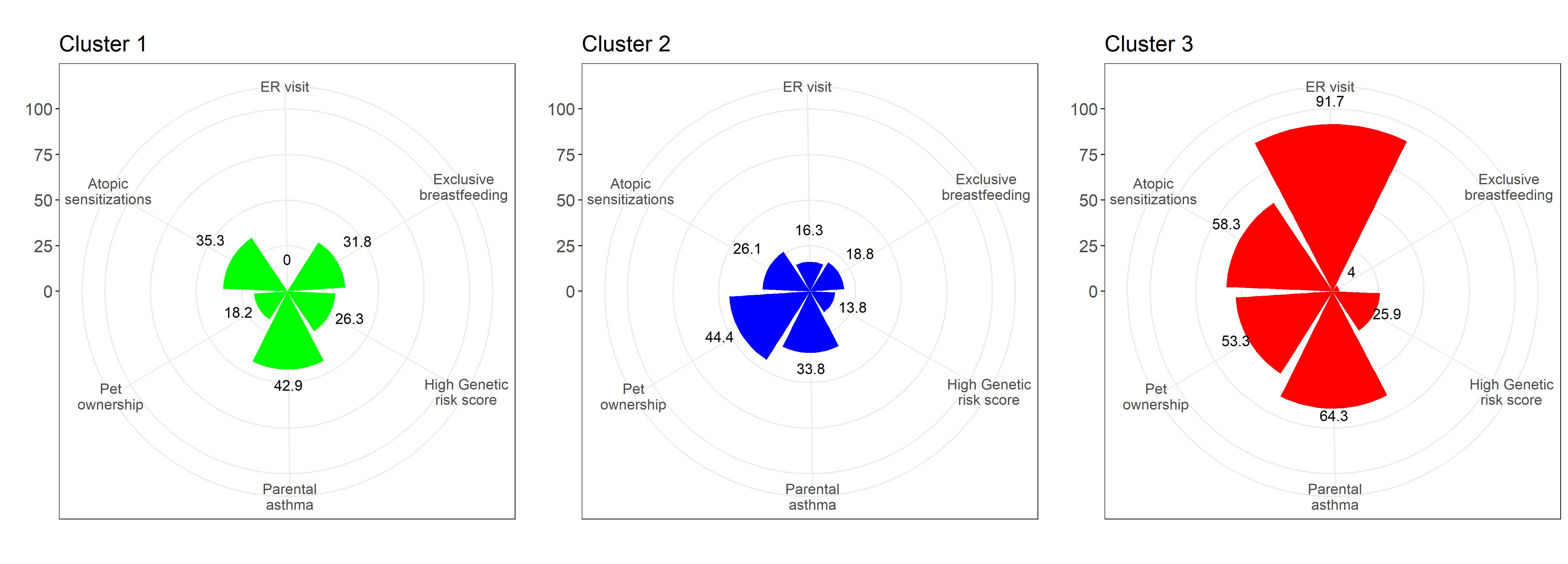}
\caption{Polor plot for the percentage of gene and environment exposure variables by global clustering $\bm{C}$. Variables are emergency room (ER) visits during the first 5 years, exclusive breastfeeding at 6 months, high genetic risk score (defined as genetic risk score $>$ 1), parental asthma status, pet ownership status,  atopic sensitization at 5 years.}
\end{figure}

\newpage
 \begin{figure}[htbp]  
 \centering
   \includegraphics[width=11cm,height=11cm]{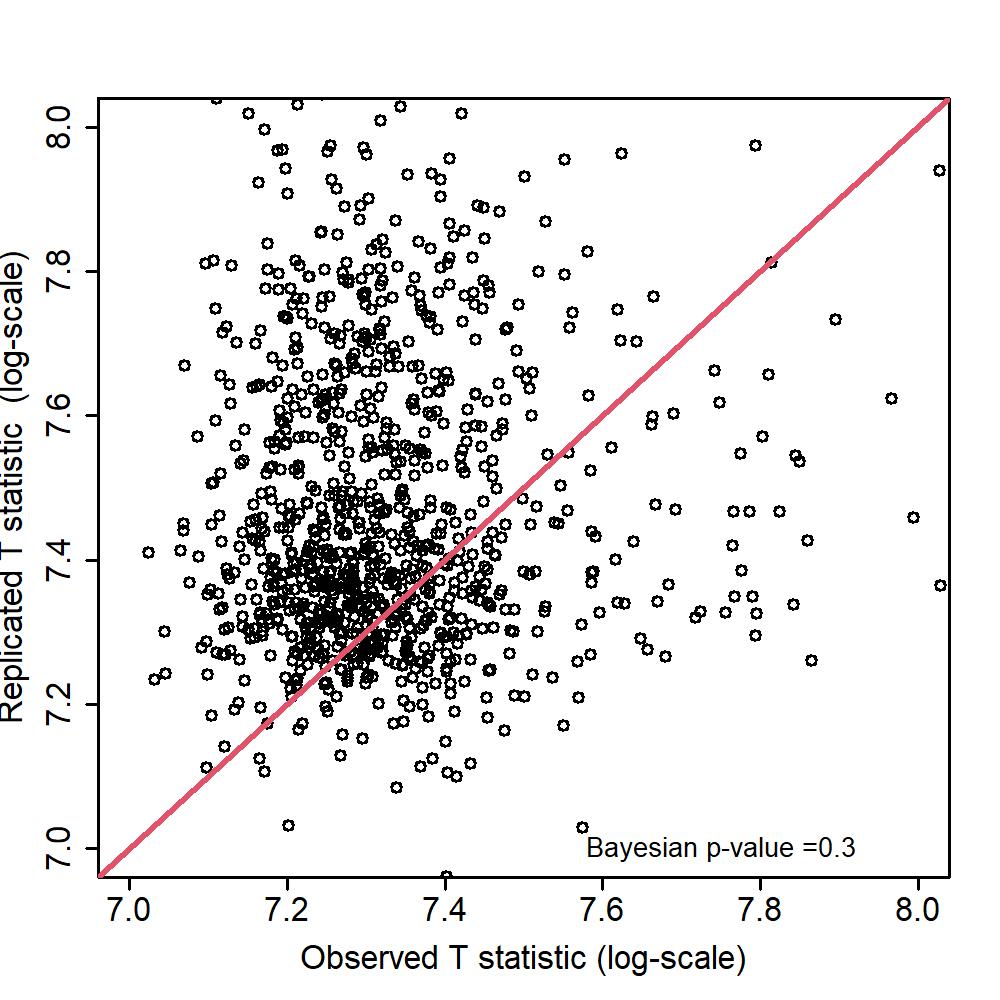}
\caption{Agreement between Replicated T  statistic and Observed T statistic and Bayesian P value.}
\end{figure}

 \newpage
\begin{figure}[htbp]  
 \centering
   \includegraphics[width=11cm,height=13cm]{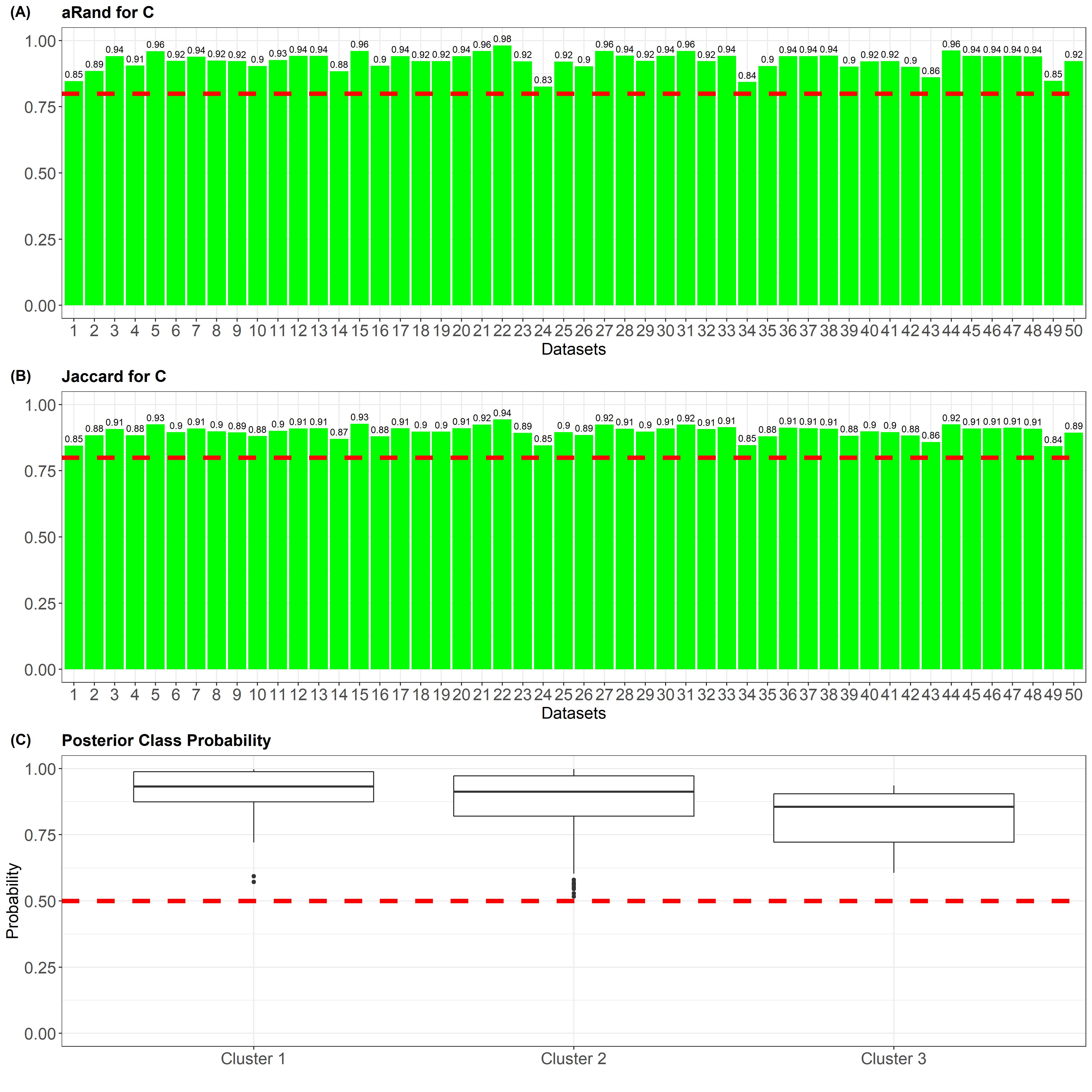}
\caption{Cluster stability of the three-class model. (A) aRand for global clustering $\bm{C}$ based on 50 subsets of the original data. Reference line represents 0.8. (B) Jaccard index for global clustering $\bm{C}$ based on 50 subsets of the original data. Reference line represents 0.8. (C) Posterior class probability. The reference line represents 0.5.}
\end{figure}

\newpage
\begin{figure}[htbp]  
 \centering
   \includegraphics[width=14cm,height=5cm]{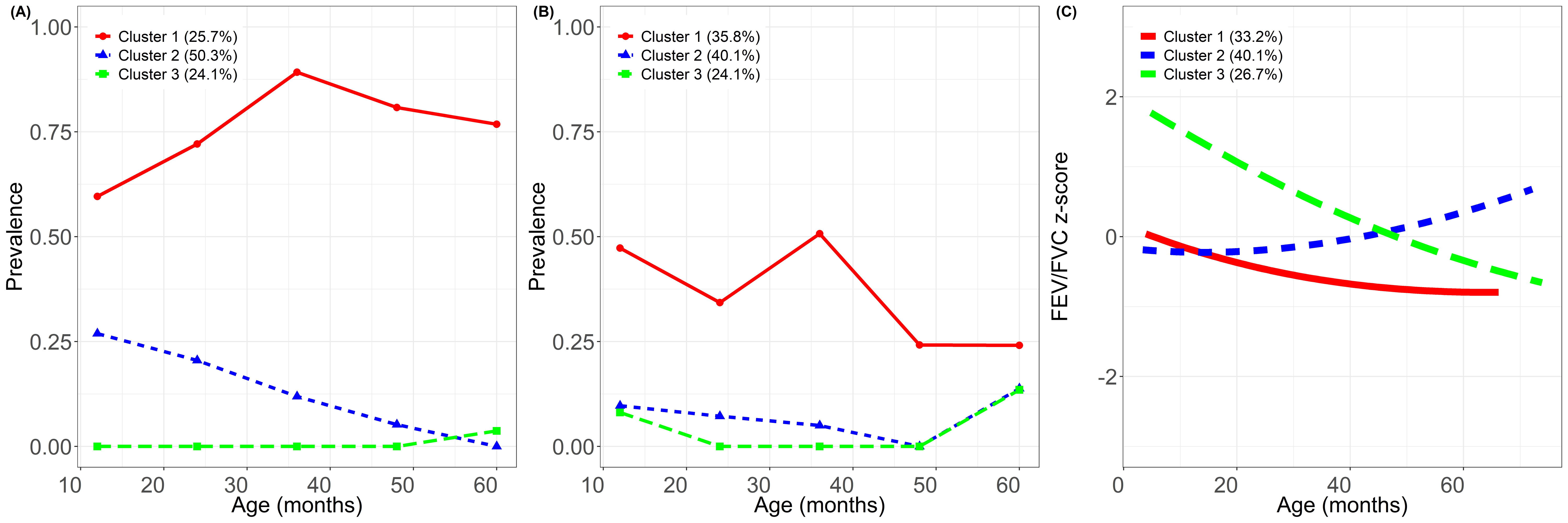}
\caption{Clustering results based on a single longitudinal marker. (A) wheeze patterns, (B) cough patterns, (C) FEV/FVC z-score patterns.}
\end{figure}

\newpage
\begin{figure}[htbp]  
 \centering
   \includegraphics[width=14cm,height=15cm]{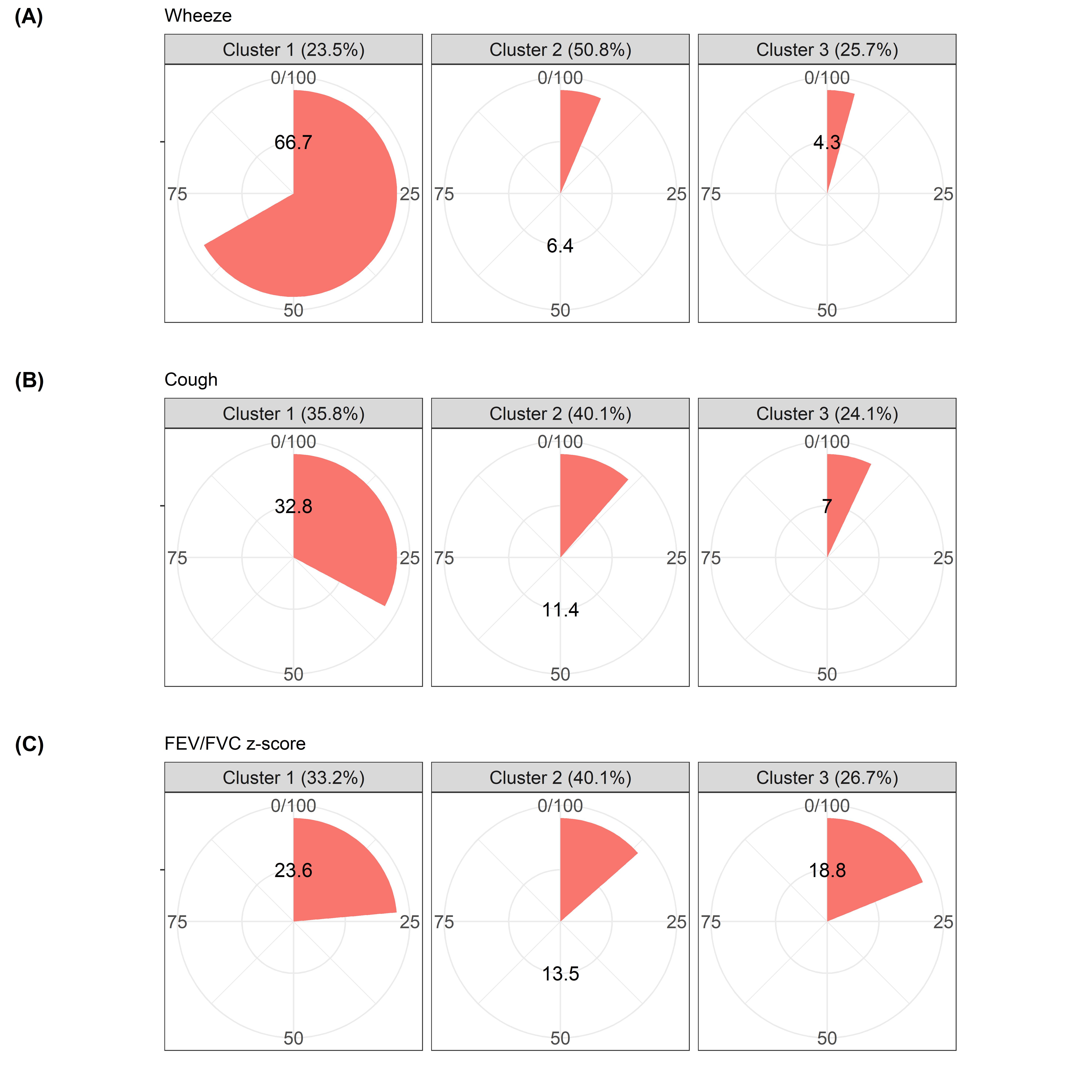}
\caption{Asthma prevalence by clusters based on separated clustering.  }
\end{figure}

\newpage
\begin{figure}[htbp]  
 \centering
   \includegraphics[width=15cm,height=11cm]{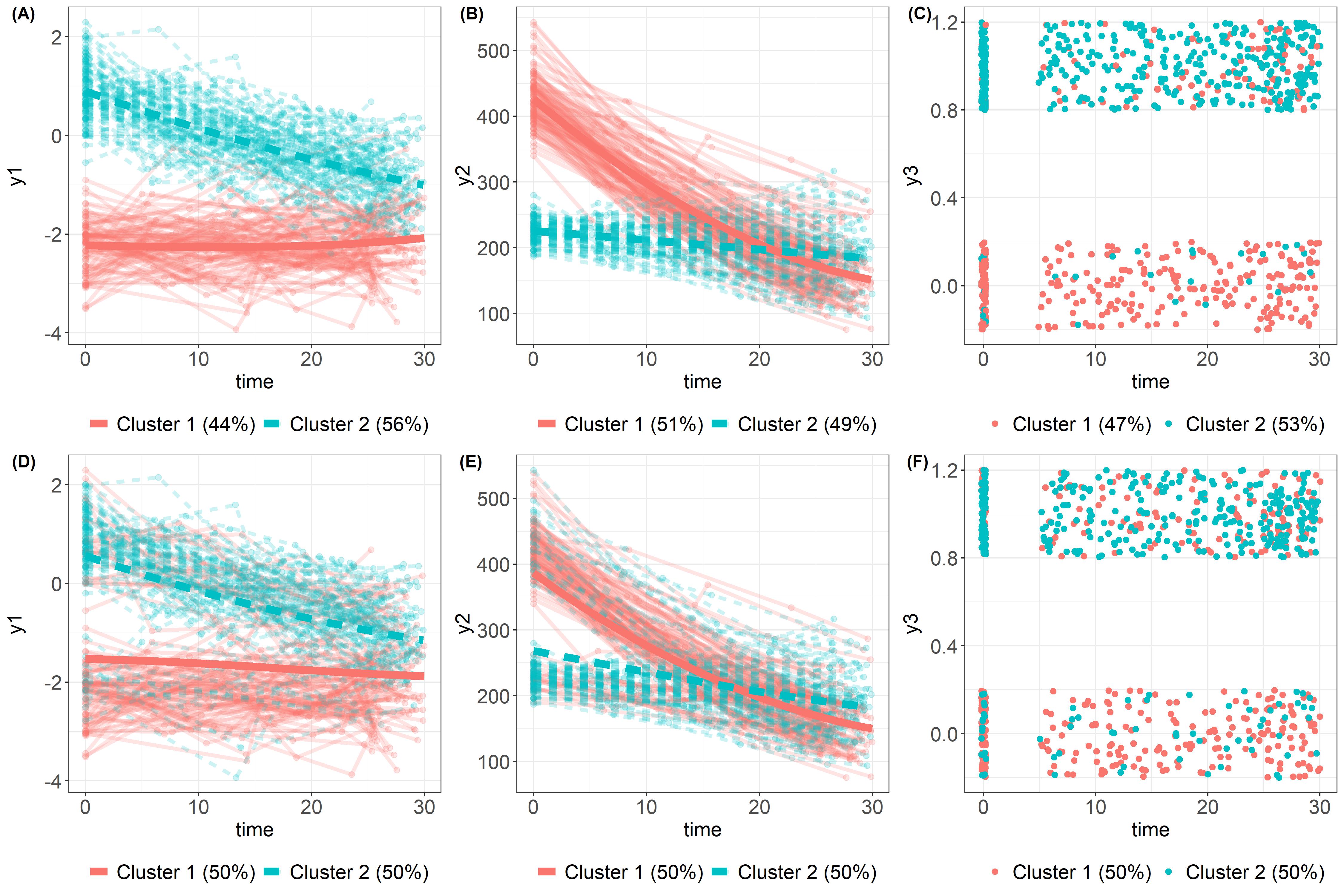}
\caption{Simulation scenario based on a two-cluster model ($K=2$) for a randomly selected simulated dataset. (A) Continuous marker plotted by true local clustering $\bm{L}_1$. (B) Count marker plotted by true local clustering $\bm{L}_2$. (C) Binary marker plotted by true local clustering $\bm{L}_3$ (jittering effect is applied for visualization purpose). (D) Continuous marker plotted by true global clustering $\bm{C}$. (E) Count marker plotted by true global clustering $\bm{C}$. (F) Binary marker plotted by true global clustering $\bm{C}$ (jittering effect is applied for visualization purpose).  }
\end{figure}

\newpage
\begin{figure}[htbp]  
 \centering
   \includegraphics[width=15cm,height=11cm]{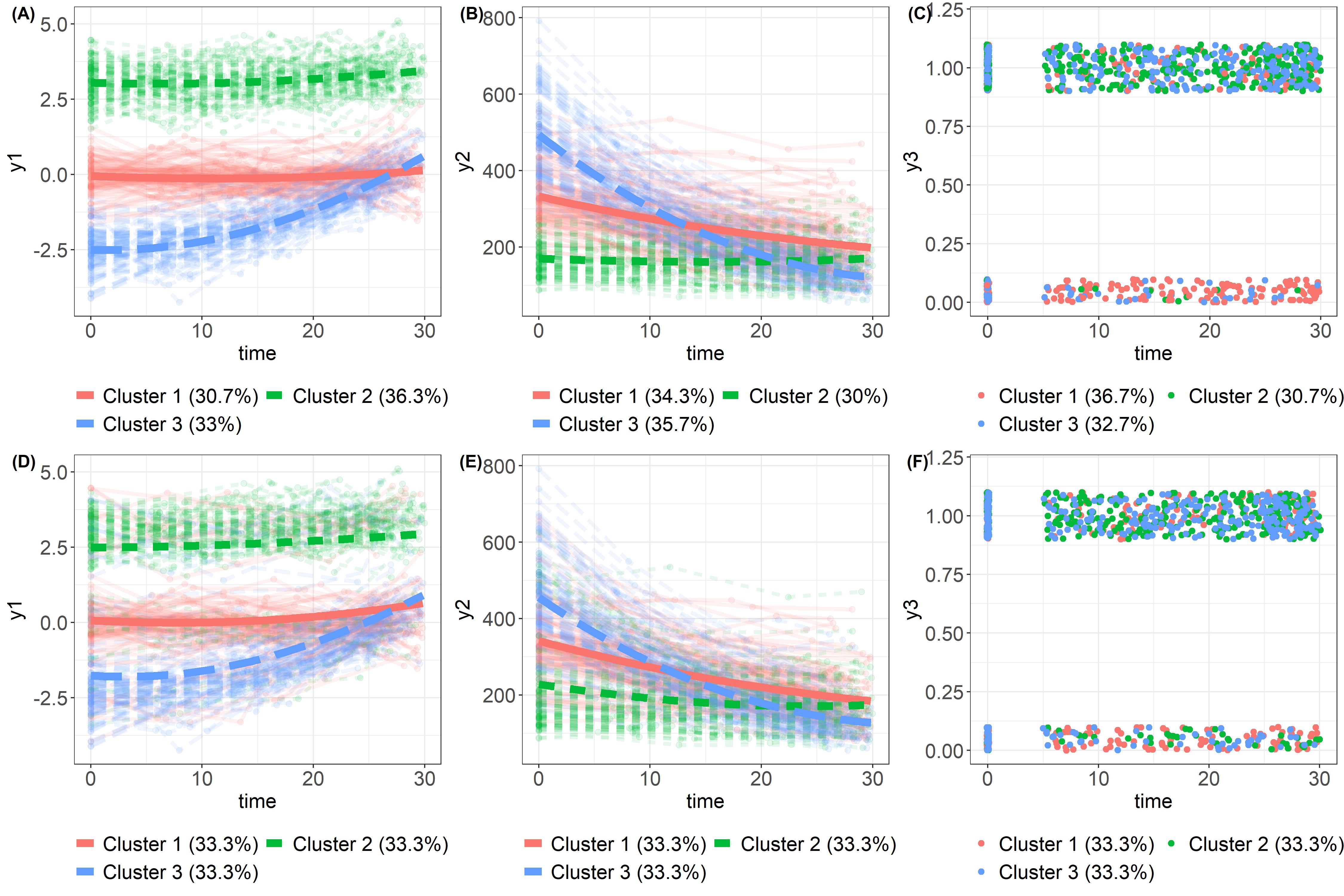}
\caption{Simulation scenario based on a three-cluster model ($K=3$)  for a randomly selected simulated dataset. (A) Continuous marker plotted by true local clustering $\bm{L}_1$. (B) Count marker plotted by true local clustering $\bm{L}_2$. (C) Binary marker plotted by true local clustering $\bm{L}_3$ (jittering effect is applied for visualization purpose). (D) Continuous marker plotted by true global clustering $\bm{C}$. (E) Count marker plotted by true global clustering $\bm{C}$. (F) Binary marker plotted by true global clustering $\bm{C}$ (jittering effect is applied for visualization purpose).   }
\end{figure}

\newpage
\begin{figure}[htbp]  
 \centering
   \includegraphics[width=15cm,height=11cm]{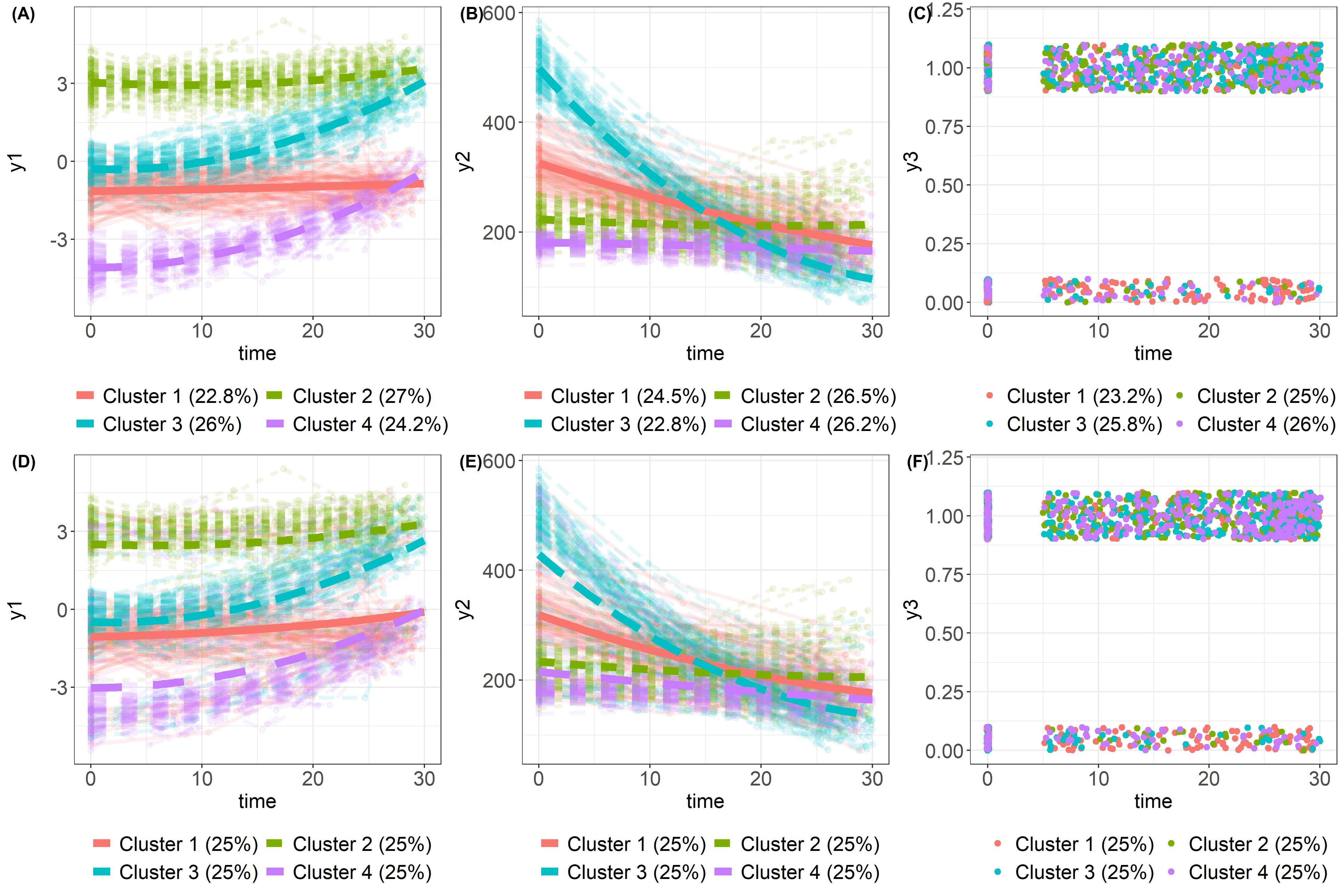}
\caption{Simulation scenario based on a four-cluster model ($K=4$)  for a randomly selected simulated dataset. (A) Continuous marker plotted by true local clustering $\bm{L}_1$. (B) Count marker plotted by true local clustering $\bm{L}_2$. (C) Binary marker plotted by true local clustering $\bm{L}_3$ (jittering effect is applied for visualization purpose). (D) Continuous marker plotted by true global clustering $\bm{C}$. (E) Count marker plotted by true global clustering $\bm{C}$. (F) Binary marker plotted by true global clustering $\bm{C}$ (jittering effect is applied for visualization purpose).   }
\end{figure}

\newpage
\begin{table}[htbp]
  \centering
  \caption{Demographic and clinical information}
  \scalebox{1}{  \begin{tabular}{lc}
    \toprule
                      & \multicolumn{1}{p{14.5em}}{\textbf{Overall}} \\
                      & \multicolumn{1}{p{14.5em}}{\textbf{(N=187)}} \\
    \midrule
    \textbf{Sex}      &  \\
    m                 & 111 (59.4\%) \\
    f                 & 76 (40.6\%) \\
    \textbf{Gestational age (wk)} &  \\
    Mean (SD)         & 39.1 (1.24) \\
    \textbf{Mother Education} &  \\
    High-school or below & 7 (3.7\%) \\
    College to university & 121 (64.7\%) \\
    Above unviersity  & 32 (17.1\%) \\
    Missing           & 27 (14.4\%) \\
    \textbf{Father Education} &  \\
    High-school or below & 14 (7.5\%) \\
    College to university & 115 (61.5\%) \\
    Above unviersity  & 26 (13.9\%) \\
    Missing           & 32 (17.1\%) \\
    \textbf{Income}   &  \\
    $<$40,000           & 5 (2.7\%) \\
    \$40,000 - \$79,000 & 21 (11.2\%) \\
    \$80,000 - \$149.9999 & 63 (33.7\%) \\
    \$150,000 or over & 58 (31.0\%) \\
    prefer not to say & 13 (7.0\%) \\
    Missing           & 27 (14.4\%) \\
    \textbf{Birth weight z-score} &  \\
    Mean (SD)         & 0.262 (0.935) \\
    \textbf{Asthma at 5 years} &  \\
    No                &  155 (82.9\%) \\
    Definite/Possible & 32 (17.1\%) \\
    \bottomrule
    \end{tabular}}
\end{table}

\newpage

\begin{table}[htbp]
  \centering
  \caption{Comparison of clustering results between model with random intercept and model with random intercept and slope for local clusterings and global clustering}
   \scalebox{0.8}{    \begin{tabular}{ccccc}
    \toprule
          &       & \multicolumn{3}{c}{\textbf{Model with random intercept}} \\
\cmidrule{2-5}          & \multicolumn{1}{p{8em}}{\textbf{wheeze-specific clustering}} & Cluster 1 & Cluster 2 & Cluster 3 \\
    \multicolumn{1}{c}{\multirow{3}[0]{*}{\textbf{Model with random intercept and slope}}} & Cluster 1 & 33    & 1     & 1 \\
          & Cluster 2 & 0     & 39    & 0 \\
          & Cluster 3 & 0     & 2     & 111 \\
          &       & \multicolumn{3}{c}{\textbf{Model with random intercept}} \\
          & \multicolumn{1}{p{8em}}{\textbf{cough-specific clustering}} & Cluster 1 & Cluster 2 & Cluster 3 \\
    \multicolumn{1}{c}{\multirow{3}[0]{*}{\textbf{Model with random  intercept and slope}}} & Cluster 1 & 31    & 2     & 1 \\
          & Cluster 2 & 0     & 37    & 0 \\
          & Cluster 3 & 1     & 1     & 114 \\
          &       & \multicolumn{3}{c}{\textbf{Model with random intercept}} \\
          & \multicolumn{1}{p{8em}}{\textbf{FEVFVC-specific clustering}} & Cluster 1 & Cluster 2 & Cluster 3 \\
    \multicolumn{1}{c}{\multirow{3}[0]{*}{\textbf{Model with random intercept and slope}}} & Cluster 1 & 38    & 0     & 0 \\
          & Cluster 2 & 0     & 34    & 1 \\
          & Cluster 3 & 1     & 2     & 111 \\
          &       & \multicolumn{3}{c}{\textbf{Model with random intercept}} \\
          & \textbf{Global clustering} & Cluster 1 & Cluster 2 & Cluster 3 \\
    \multicolumn{1}{c}{\multirow{3}[1]{*}{\textbf{Model with random intercept and slope}}} & Cluster 1 & 19    & 0     & 1 \\
          & Cluster 2 & 0     & 29    & 0 \\
          & Cluster 3 & 0     & 2     & 136 \\
    \bottomrule
    \end{tabular}}%
  \label{tab:addlabel}%
\end{table}%

\newpage
\begin{table}[htbp]
  \centering
  \caption{Cluster proportions under different hyperparameters of priors of $\bm{\pi}$ and $\bm{\alpha}$}
  \scalebox{0.8} {\begin{tabular}{ccccrrrrrrrrrrrr}
    \toprule
                     &                   &                   &                   & \multicolumn{3}{c}{$\bm{C}$}                              & \multicolumn{3}{c}{$\bm{L}_1$}                            & \multicolumn{3}{c}{$\bm{L}_2$}                            & \multicolumn{3}{c}{$\bm{L}_3$} \\
\cmidrule{2-16}                      & $\phi$            & $\delta_1$        & $\delta_2$        & \multicolumn{1}{c}{$k=1$} & \multicolumn{1}{c}{$k=2$} & \multicolumn{1}{c}{$k=3$} & \multicolumn{1}{c}{$k=1$} & \multicolumn{1}{c}{$k=2$} & \multicolumn{1}{c}{$k=3$} & \multicolumn{1}{c}{$k=1$} & \multicolumn{1}{c}{$k=2$} & \multicolumn{1}{c}{$k=3$} & \multicolumn{1}{c}{$k=1$} & \multicolumn{1}{c}{$k=2$} & \multicolumn{1}{c}{$k=3$} \\
    \midrule
    Prior 1           & 1                 & 1                 & 1                 & 16                & 73.3              & 10.7              & 21.4              & 58.3              & 20.3              & 20.9              & 58.8              & 20.3              & 19.3              & 60.4              & 20.3 \\
    Prior 2           & 1                 & 8                 & 3                 & 16.6              & 69                & 14.4              & 22.5              & 51.9              & 25.7              & 20.9              & 52.9              & 26.2              & 19.8              & 57.2              & 23 \\
    Prior 3           & 1                 & 1                 & 5                 & 16.6              & 69                & 14.4              & 22.5              & 51.9              & 25.7              & 20.9              & 52.9              & 26.2              & 19.8              & 57.2              & 23 \\
    Prior 4           & 1                 & 5                 & 1                 & 13.4              & 54                & 32.6              & 19.3              & 41.2              & 39.6              & 16.6              & 40.6              & 42.8              & 17.1              & 48.7              & 34.2 \\
    Prior 5           & 5                 & 1                 & 1                 & 19.8              & 65.8              & 14.4              & 23                & 53.5              & 23.5              & 24.1              & 46.5              & 29.4              & 22.5              & 54.5              & 23 \\
    Prior 6           & 5                 & 8                 & 3                 & 19.3              & 43.9              & 36.9              & 21.9              & 35.3              & 42.8              & 21.9              & 32.1              & 46                & 20.9              & 43.3              & 35.8 \\
    Prior 7           & 5                 & 1                 & 5                 & 12.8              & 3.2               & 84                & 19.3              & 2.7               & 78.1              & 15                & 13.9              & 71.1              & 21.9              & 20.9              & 57.2 \\
    Prior 8           & 5                 & 5                 & 1                 & 17.6              & 17.6              & 64.7              & 22.5              & 33.2              & 44.4              & 20.9              & 46                & 33.2              & 25.7              & 24.6              & 49.7 \\
    Prior 9           & 10                & 1                 & 1                 & 19.8              & 49.2              & 31                & 23                & 38.5              & 38.5              & 24.1              & 32.1              & 43.9              & 24.1              & 47.6              & 28.3 \\
    Prior 10          & 10                & 8                 & 3                 & 17.6              & 69                & 13.4              & 21.9              & 61.5              & 16.6              & 22.5              & 59.4              & 18.2              & 18.7              & 61                & 20.3 \\
    Prior 11          & 10                & 1                 & 5                 & 16.6              & 53.5              & 29.9              & 21.4              & 43.3              & 35.3              & 19.8              & 39                & 41.2              & 20.3              & 50.3              & 29.4 \\
    Prior 12          & 10                & 5                 & 1                 & 20.3              & 60.4              & 19.3              & 22.5              & 46.5              & 31                & 24.6              & 42.2              & 33.2              & 20.9              & 50.8              & 28.3 \\
    \bottomrule
    \end{tabular}}
\end{table}%

\newpage
\begin{table}[htbp]
  \centering
  \caption{Root mean square error of parameter estimates for model with $K=2$ under random effects with normal distribution}
    \scalebox{0.8}{   \begin{threeparttable}
     \begin{tabular}{lcccccccc}
    \toprule
                      & \multicolumn{2}{c}{$\bm{\alpha} =  (0.5, 0.5, 0.5) $} & \multicolumn{2}{c}{$\bm{\alpha}  =  (0.8, 0.8, 0.8) $} & \multicolumn{2}{c}{$\bm{\alpha} =  (1, 1, 1) $} & \multicolumn{2}{c}{$\bm{\alpha} =  (1, 0.5, 0.8) $} \\
    \midrule
                      & $k=1$             & $k=2$             & $k=1$             & $k=2$             & $k=1$             & $k=2$             & $k=1$             & $k=2$ \\
    \midrule
    $\pi_k$           & 0.138             & 0.138             & 0.030             & 0.030             & 0.003             & 0.003             & 0.020             & 0.020 \\
    $r = 1$ (Continous) &                   &                   &                   &                   &                   &                   &                   &  \\
    $\alpha_r$        & \multicolumn{2}{c}{0.108}             & \multicolumn{2}{c}{0.051}             & \multicolumn{2}{c}{0.005}             & \multicolumn{2}{c}{0.127} \\
    $\gamma_{1k,r}$   & 0.050             & 0.072             & 0.056             & 0.052             & 0.056             & 0.059             & 0.054             & 0.059 \\
    $\gamma_{2k,r}$   & 0.010             & 0.011             & 0.010             & 0.011             & 0.012             & 0.011             & 0.011             & 0.011 \\
    $\gamma_{3k,r}$   & 0.000             & 0.000             & 0.000             & 0.000             & 0.000             & 0.000             & 0.000             & 0.000 \\
    $\Sigma_{11k,r}$  & 0.000             & 0.000             & 0.000             & 0.000             & 0.000             & 0.000             & 0.000             & 0.000 \\
    $\Sigma_{22k,r}$  & 0.000             & 0.000             & 0.000             & 0.000             & 0.000             & 0.000             & 0.000             & 0.000 \\
    $r = 2$  (Count)  &                   &                   &                   &                   &                   &                   &                   &  \\
    $\alpha_r$        & \multicolumn{2}{c}{0.118}             & \multicolumn{2}{c}{0.062}             & \multicolumn{2}{c}{0.006}             & \multicolumn{2}{c}{0.045} \\
    $\gamma_{1k,r}$   & 0.618             & 0.618             & 0.621             & 0.618             & 0.619             & 0.617             & 0.619             & 0.619 \\
    $\gamma_{2k,r}$   & 0.029             & 0.027             & 0.029             & 0.028             & 0.029             & 0.027             & 0.029             & 0.027 \\
    $\Sigma_{11k,r}$  & 0.006             & 0.001             & 0.006             & 0.001             & 0.006             & 0.001             & 0.006             & 0.001 \\
    $\Sigma_{22k,r}$  & 0.000             & 0.000             & 0.000             & 0.000             & 0.000             & 0.000             & 0.000             & 0.000 \\
    $r = 3$ (Binary)  &                   &                   &                   &                   &                   &                   &                   &  \\
    $\alpha_r$        & \multicolumn{2}{c}{0.120}             & \multicolumn{2}{c}{0.050}             & \multicolumn{2}{c}{0.008}             & \multicolumn{2}{c}{0.099} \\
    $\gamma_{1k,r}$   & 0.317             & 0.317             & 0.219             & 0.344             & 0.220             & 0.288             & 0.215             & 0.346 \\
    $\gamma_{2k,r}$   & 0.015             & 0.015             & 0.012             & 0.024             & 0.012             & 0.020             & 0.011             & 0.023 \\
    $\Sigma_{11k,r}$  & 0.000             & 0.000             & 0.000             & 0.000             & 0.000             & 0.000             & 0.000             & 0.000 \\
    \bottomrule
    \end{tabular}    
    \begin{tablenotes} 
       \item $\Sigma_{11k,r}$ and $\Sigma_{22k,r}$ denote the diagonal elements of $\Sigma_{k,r}$
	\end{tablenotes}	  
\end{threeparttable}}
\end{table}%

\newpage
\begin{table}[htbp]
  \centering
  \caption{Root mean square error of parameter estimates for model with $K=2$ under random effects with $t$ distribution}  
  \scalebox{0.8} {\begin{threeparttable} 
  \begin{tabular}{lcccccccc}
    \toprule
                      & \multicolumn{2}{l}{$\bm{\alpha} =  (0.5, 0.5, 0.5) $} & \multicolumn{2}{l}{$\bm{\alpha}  =  (0.8, 0.8, 0.8) $} & \multicolumn{2}{l}{$\bm{\alpha} =  (1, 1, 1) $} & \multicolumn{2}{l}{$\bm{\alpha} =  (1, 0.5, 0.8) $} \\
    \midrule
                      & $k=1$             & $k=2$             & $k=1$             & $k=2$             & $k=1$             & $k=2$             & $k=1$             & $k=2$ \\
    \midrule
    $\pi_k$           & 0.143             & 0.143             & 0.028             & 0.028             & 0.003             & 0.003             & 0.021             & 0.021 \\
    $r = 1$ (Continous) &                   &                   &                   &                   &                   &                   &                   &  \\
    $\alpha_r$        & \multicolumn{2}{c}{0.112}             & \multicolumn{2}{c}{0.044}             & \multicolumn{2}{c}{0.005}             & \multicolumn{2}{c}{0.122} \\
    $\gamma_{1k,r}$   & 0.063             & 0.062             & 0.069             & 0.067             & 0.068             & 0.058             & 0.058             & 0.065 \\
    $\gamma_{2k,r}$   & 0.011             & 0.011             & 0.011             & 0.010             & 0.012             & 0.013             & 0.011             & 0.013 \\
    $\gamma_{3k,r}$   & 0.000             & 0.000             & 0.000             & 0.000             & 0.000             & 0.000             & 0.000             & 0.000 \\
    $\Sigma_{11k,r}$  & 0.000             & 0.000             & 0.000             & 0.000             & 0.000             & 0.000             & 0.000             & 0.000 \\
    $\Sigma_{22k,r}$  & 0.000             & 0.000             & 0.000             & 0.000             & 0.000             & 0.000             & 0.000             & 0.000 \\
    $r = 2$  (Count)  &                   &                   &                   &                   &                   &                   &                   &  \\
    $\alpha_r$        & \multicolumn{2}{c}{0.114}             & \multicolumn{2}{c}{0.050}             & \multicolumn{2}{c}{0.011}             & \multicolumn{2}{c}{0.041} \\
    $\gamma_{1k,r}$   & 0.617             & 0.622             & 0.618             & 0.622             & 0.619             & 0.621             & 0.617             & 0.627 \\
    $\gamma_{2k,r}$   & 0.030             & 0.027             & 0.030             & 0.027             & 0.030             & 0.027             & 0.029             & 0.027 \\
    $\Sigma_{11k,r}$  & 0.006             & 0.001             & 0.006             & 0.001             & 0.006             & 0.001             & 0.006             & 0.001 \\
    $\Sigma_{22k,r}$  & 0.000             & 0.000             & 0.000             & 0.000             & 0.000             & 0.000             & 0.000             & 0.000 \\
    $r = 3$ (Binary)  &                   &                   &                   &                   &                   &                   &                   &  \\
    $\alpha_r$        & \multicolumn{2}{c}{0.136}             & \multicolumn{2}{c}{0.052}             & \multicolumn{2}{c}{0.009}             & \multicolumn{2}{c}{0.097} \\
    $\gamma_{1k,r}$   & 0.249             & 0.410             & 0.261             & 0.381             & 0.247             & 0.370             & 0.256             & 0.392 \\
    $\gamma_{2k,r}$   & 0.012             & 0.669             & 0.014             & 0.024             & 0.014             & 0.023             & 0.012             & 0.018 \\
    $\Sigma_{11k,r}$  & 0.000             & 0.000             & 0.000             & 0.000             & 0.000             & 0.000             & 0.000             & 0.000 \\
    \bottomrule
    \end{tabular}    
    \begin{tablenotes} 
       \item $\Sigma_{11k,r}$ and $\Sigma_{22k,r}$ denote the diagonal elements of $\Sigma_{k,r}$
	\end{tablenotes}	  
\end{threeparttable}}
\end{table}%

\newpage
\KOMAoptions{paper=landscape,pagesize}

\begin{table}[htbp]
  \centering
  \caption{Root mean square error of parameter estimates for model with $K=3$ under random effects with normal distribution}  
    \begin{threeparttable} 
	\begin{tabular}{lcccccccccccc}
    \toprule
                      & \multicolumn{3}{c}{$\bm{\alpha} =  (0.34, 0.34, 0.34) $}  & \multicolumn{3}{c}{$\bm{\alpha}  =  (0.8, 0.8, 0.8) $}    & \multicolumn{3}{c}{$\bm{\alpha} =  (1, 1, 1) $}           & \multicolumn{3}{c}{$\bm{\alpha} =  (1, 0.34, 0.8) $} \\
    \midrule
                      & $k=1$             & $k=2$             & $k=3$             & $k=1$             & $k=2$             & $k=3$             & $k=1$             & $k=2$             & $k=3$             & $k=1$             & $k=2$             & $k=3$ \\
    \midrule
    $\pi_k$            & 0.148             & 0.163             & 0.210             & 0.025             & 0.018             & 0.027             & 0.003             & 0.003             & 0.004             & 0.013             & 0.012             & 0.012 \\
    $r = 1$ (Continous) &                   &                   &                   &                   &                   &                   &                   &                   &                   &                   &                   &  \\
    $\alpha_r$         & \multicolumn{3}{c}{0.108}                                 & \multicolumn{3}{c}{0.060}                                 & \multicolumn{3}{c}{0.006}                                 & \multicolumn{3}{c}{0.134} \\
    $\gamma_{1k,r}$   & 0.059             & 0.068             & 0.064             & 0.061             & 0.051             & 0.062             & 0.072             & 0.060             & 0.062             & 0.057             & 0.072             & 0.056 \\
    $\gamma_{2k,r}$   & 0.011             & 0.011             & 0.011             & 0.011             & 0.012             & 0.011             & 0.012             & 0.010             & 0.010             & 0.010             & 0.012             & 0.011 \\
    $\gamma_{3k,r}$   & 0.000             & 0.000             & 0.000             & 0.000             & 0.000             & 0.000             & 0.000             & 0.000             & 0.000             & 0.000             & 0.000             & 0.000 \\
    $\Sigma_{11k,r}$  & 0.000             & 0.000             & 0.000             & 0.000             & 0.000             & 0.000             & 0.000             & 0.000             & 0.000             & 0.000             & 0.000             & 0.000 \\
    $\Sigma_{22k,r}$  & 0.000             & 0.000             & 0.000             & 0.000             & 0.000             & 0.000             & 0.000             & 0.000             & 0.000             & 0.000             & 0.000             & 0.000 \\
    $r = 2$  (Count)  &                   &                   &                   &                   &                   &                   &                   &                   &                   &                   &                   &  \\
    $\alpha_r$         & \multicolumn{3}{c}{0.127}                                 & \multicolumn{3}{c}{0.132}                                 & \multicolumn{3}{c}{0.118}                                 & \multicolumn{3}{c}{0.036} \\
    $\gamma_{1k,r}$   & 0.398             & 0.167             & 0.397             & 0.391             & 0.057             & 0.398             & 0.390             & 0.061             & 0.376             & 0.390             & 0.068             & 0.398 \\
    $\gamma_{2k,r}$   & 0.028             & 0.009             & 0.038             & 0.028             & 0.006             & 0.037             & 0.028             & 0.004             & 0.036             & 0.028             & 0.007             & 0.038 \\
    $\Sigma_{11k,r}$  & 0.036             & 0.052             & 0.016             & 0.037             & 0.052             & 0.016             & 0.037             & 0.057             & 0.020             & 0.036             & 0.052             & 0.016 \\
    $\Sigma_{22k,r}$  & 0.000             & 0.000             & 0.000             & 0.000             & 0.000             & 0.000             & 0.000             & 0.000             & 0.000             & 0.000             & 0.000             & 0.000 \\
    $r = 3$ (Binary)  &                   &                   &                   &                   &                   &                   &                   &                   &                   & \multicolumn{2}{c}{}                  &  \\
    $\alpha_r$         & \multicolumn{3}{c}{0.219}                                 & \multicolumn{3}{c}{0.075}                                 & \multicolumn{3}{c}{0.014}                                 & \multicolumn{3}{c}{0.104} \\
    $\gamma_{1k,r}$   & 0.485             & 4.914             & 1.076             & 0.269             & 2.973             & 0.253             & 0.223             & 0.348             & 0.192             & 0.237             & 0.490             & 0.234 \\
    $\gamma_{2k,r}$   & 0.021             & 5.969             & 1.221             & 0.014             & 3.080             & 0.014             & 0.011             & 0.022             & 0.013             & 0.013             & 0.026             & 0.017 \\
    $\Sigma_{11k,r}$  & 0.000             & 0.000             & 0.000             & 0.000             & 0.000             & 0.000             & 0.000             & 0.000             & 0.000             & 0.000             & 0.000             & 0.000 \\
    \bottomrule
    \end{tabular}    
    \begin{tablenotes} 
       \item $\Sigma_{11k,r}$ and $\Sigma_{22k,r}$ denote the diagonal elements of $\Sigma_{k,r}$
	\end{tablenotes}	  
\end{threeparttable}
\end{table}%

\newpage 
\begin{table}[htbp]
  \centering
  \caption{Root mean square error of parameter estimates for model with $K=3$ under random effects with $t$ distribution}  
    \begin{threeparttable} 
    \begin{tabular}{lcccccccccccc}
    \toprule
                      & \multicolumn{3}{c}{$\bm{\alpha} =  (0.34, 0.34, 0.34) $}  & \multicolumn{3}{c}{$\bm{\alpha}  =  (0.8, 0.8, 0.8) $}    & \multicolumn{3}{c}{$\bm{\alpha} =  (1, 1, 1) $}           & \multicolumn{3}{c}{$\bm{\alpha} =  (1, 0.34, 0.8) $} \\
    \midrule
                      & $k=1$             & $k=2$             & $k=3$             & $k=1$             & $k=2$             & $k=3$             & $k=1$             & $k=2$             & $k=3$             & $k=1$             & $k=2$             & $k=3$ \\
    \midrule
    $\pi_k$            & 0.168             & 0.164             & 0.178             & 0.036             & 0.024             & 0.030             & 0.006             & 0.005             & 0.006             & 0.016             & 0.017             & 0.015 \\
    $r = 1$ (Continous) &                   &                   &                   &                   &                   &                   &                   &                   &                   &                   &                   &  \\
    $\alpha_r$         & \multicolumn{3}{c}{0.088}                                 & \multicolumn{3}{c}{0.047}                                 & \multicolumn{3}{c}{0.005}                                 & \multicolumn{3}{c}{0.141} \\
    $\gamma_{1k,r}$   & 0.058             & 0.056             & 0.071             & 0.064             & 0.055             & 0.063             & 0.058             & 0.053             & 0.057             & 0.061             & 0.070             & 0.054 \\
    $\gamma_{2k,r}$   & 0.011             & 0.012             & 0.011             & 0.011             & 0.014             & 0.009             & 0.011             & 0.011             & 0.010             & 0.011             & 0.014             & 0.011 \\
    $\gamma_{3k,r}$   & 0.000             & 0.000             & 0.000             & 0.000             & 0.000             & 0.000             & 0.000             & 0.000             & 0.000             & 0.000             & 0.000             & 0.000 \\
    $\Sigma_{11k,r}$  & 0.000             & 0.000             & 0.000             & 0.000             & 0.000             & 0.000             & 0.000             & 0.000             & 0.000             & 0.000             & 0.000             & 0.000 \\
    $\Sigma_{22k,r}$  & 0.000             & 0.000             & 0.000             & 0.000             & 0.000             & 0.000             & 0.000             & 0.000             & 0.000             & 0.000             & 0.000             & 0.000 \\
    $r = 2$  (Count)  &                   &                   &                   &                   &                   &                   &                   &                   &                   &                   &                   &  \\
    $\alpha_r$         & \multicolumn{3}{c}{0.137}                                 & \multicolumn{3}{c}{0.163}                                 & \multicolumn{3}{c}{0.205}                                 & \multicolumn{3}{c}{0.029} \\
    $\gamma_{1k,r}$   & 0.400             & 0.072             & 0.382             & 0.395             & 0.066             & 0.363             & 0.398             & 0.071             & 0.369             & 0.398             & 0.064             & 0.380 \\
    $\gamma_{2k,r}$   & 0.028             & 0.008             & 0.039             & 0.028             & 0.008             & 0.039             & 0.028             & 0.008             & 0.039             & 0.028             & 0.008             & 0.040 \\
    $\Sigma_{11k,r}$  & 0.035             & 0.051             & 0.015             & 0.036             & 0.052             & 0.015             & 0.036             & 0.052             & 0.015             & 0.036             & 0.052             & 0.015 \\
    $\Sigma_{22k,r}$  & 0.000             & 0.000             & 0.000             & 0.000             & 0.000             & 0.000             & 0.000             & 0.000             & 0.000             & 0.000             & 0.000             & 0.000 \\
    $r = 3$ (Binary)  &                   &                   &                   &                   &                   &                   &                   &                   &                   & \multicolumn{2}{c}{}                  &  \\
    $\alpha_r$         & \multicolumn{3}{c}{0.210}                                 & \multicolumn{3}{c}{0.061}                                 & \multicolumn{3}{c}{0.015}                                 & \multicolumn{3}{c}{0.112} \\
    $\gamma_{1k,r}$   & 1.010             & 5.050             & 0.991             & 0.403             & 3.730             & 0.292             & 0.241             & 0.360             & 0.155             & 0.269             & 1.329             & 0.166 \\
    $\gamma_{2k,r}$   & 0.030             & 4.739             & 0.031             & 0.017             & 2.810             & 0.018             & 0.013             & 0.022             & 0.010             & 0.016             & 0.052             & 0.011 \\
    $\Sigma_{11k,r}$  & 0.000             & 0.000             & 0.000             & 0.000             & 0.000             & 0.000             & 0.000             & 0.000             & 0.000             & 0.000             & 0.000             & 0.000 \\
    \bottomrule
    \end{tabular}    
    \begin{tablenotes} 
       \item $\Sigma_{11k,r}$ and $\Sigma_{22k,r}$ denote the diagonal elements of $\Sigma_{k,r}$
	\end{tablenotes}	  
\end{threeparttable}
\end{table}%

\newpage
\begin{table}[htbp]
  \centering
  \caption{Root mean square error of parameter estimates for model with $K=4$ under random effects with normal distribution}  
        \begin{threeparttable} 
        \begin{tabular}{lcccccccccccccccc}
    \toprule
                      & \multicolumn{4}{c}{$\bm{\alpha} =  (0.25, 0.25, 0.25) $}                      & \multicolumn{4}{c}{$\bm{\alpha}  =  (0.8, 0.8, 0.8) $}                        & \multicolumn{4}{c}{$\bm{\alpha} =  (1, 1, 1) $}                               & \multicolumn{4}{c}{$\bm{\alpha} =  (1, 0.25, 0.8) $} \\
    \midrule
                      & $k=1$             & $k=2$             & $k=3$             & $k=4$             & $k=1$             & $k=2$             & $k=3$             & $k=4$             & $k=1$             & $k=2$             & $k=3$             & $k=4$             & $k=1$             & $k=2$             & $k=3$             & $k=4$ \\
    \midrule
    $\pi_k$           & 0.116             & 0.172             & 0.116             & 0.253             & 0.014             & 0.053             & 0.016             & 0.059             & 0.003             & 0.006             & 0.002             & 0.007             & 0.008             & 0.010             & 0.009             & 0.011 \\
    $r = 1$ (Continous) &                   &                   &                   &                   &                   &                   &                   &                   &                   &                   &                   &                   &                   &                   &                   &  \\
    $\alpha_r$        & \multicolumn{4}{c}{0.058}                                                     & \multicolumn{4}{c}{0.072}                                                     & \multicolumn{4}{c}{0.011}                                                     & \multicolumn{4}{c}{0.146} \\
    $\gamma_{1k,r}$   & 0.061             & 0.066             & 0.062             & 0.064             & 0.064             & 0.068             & 0.065             & 0.061             & 0.063             & 0.062             & 0.062             & 0.062             & 0.055             & 0.074             & 0.069             & 0.052 \\
    $\gamma_{2k,r}$   & 0.010             & 0.013             & 0.011             & 0.009             & 0.012             & 0.012             & 0.012             & 0.009             & 0.010             & 0.011             & 0.010             & 0.011             & 0.009             & 0.012             & 0.011             & 0.008 \\
    $\gamma_{3k,r}$   & 0.000             & 0.000             & 0.000             & 0.000             & 0.000             & 0.000             & 0.000             & 0.000             & 0.000             & 0.000             & 0.000             & 0.000             & 0.000             & 0.000             & 0.000             & 0.000 \\
    $\Sigma_{11k,r}$  & 0.000             & 0.000             & 0.000             & 0.000             & 0.000             & 0.000             & 0.000             & 0.000             & 0.000             & 0.000             & 0.000             & 0.000             & 0.000             & 0.000             & 0.000             & 0.000 \\
    $\Sigma_{22k,r}$  & 0.000             & 0.000             & 0.000             & 0.000             & 0.000             & 0.000             & 0.000             & 0.000             & 0.000             & 0.000             & 0.000             & 0.000             & 0.000             & 0.000             & 0.000             & 0.000 \\
    $r = 2$  (Count)  &                   &                   &                   &                   &                   &                   &                   &                   &                   &                   &                   &                   &                   &                   &                   &  \\
    $\alpha_r$        & \multicolumn{4}{c}{0.213}                                                     & \multicolumn{4}{c}{0.083}                                                     & \multicolumn{4}{c}{0.102}                                                     & \multicolumn{4}{c}{0.026} \\
    $\gamma_{1k,r}$   & 0.533             & 0.770             & 0.760             & 0.628             & 0.539             & 0.770             & 0.759             & 0.625             & 0.550             & 0.772             & 0.749             & 0.618             & 0.533             & 0.771             & 0.764             & 0.622 \\
    $\gamma_{2k,r}$   & 0.012             & 0.048             & 0.052             & 0.014             & 0.012             & 0.047             & 0.052             & 0.015             & 0.013             & 0.047             & 0.051             & 0.014             & 0.012             & 0.047             & 0.053             & 0.014 \\
    $\Sigma_{11k,r}$  & 0.011             & 0.011             & 0.004             & 0.001             & 0.011             & 0.011             & 0.004             & 0.001             & 0.011             & 0.011             & 0.004             & 0.001             & 0.011             & 0.011             & 0.004             & 0.001 \\
    $\Sigma_{22k,r}$  & 0.000             & 0.000             & 0.000             & 0.000             & 0.000             & 0.000             & 0.000             & 0.000             & 0.000             & 0.000             & 0.000             & 0.000             & 0.000             & 0.000             & 0.000             & 0.000 \\
    $r = 3$ (Binary)  &                   &                   &                   &                   &                   &                   &                   &                   &                   &                   &                   &                   &                   &                   &                   &  \\
    $\alpha_r$        & \multicolumn{4}{c}{0.223}                                                     & \multicolumn{4}{c}{0.050}                                                     & \multicolumn{4}{c}{0.019}                                                     & \multicolumn{4}{c}{0.114} \\
    $\gamma_{1k,r}$   & 0.431             & 6.903             & 2.520             & 1.546             & 0.354             & 4.192             & 0.230             & 0.281             & 0.233             & 0.340             & 0.167             & 0.180             & 0.258            & 0.576            & 0.233            & 0.265 \\
    $\gamma_{2k,r}$   & 0.020             & 4.426             & 1.797             & 2.084             & 0.017             & 4.021             & 0.016             & 0.016             & 0.012             & 0.022             & 0.011             & 0.011             & 0.015            & 0.033            & 0.015            & 0.017 \\
    $\Sigma_{11k,r}$  & 0.000             & 0.000             & 0.000             & 0.000             & 0.000             & 0.000             & 0.000             & 0.000             & 0.000             & 0.000             & 0.000             & 0.000             & 0.000             & 0.000             & 0.000             & 0.000 \\
    \bottomrule
    \end{tabular}    
    \begin{tablenotes} 
       \item $\Sigma_{11k,r}$ and $\Sigma_{22k,r}$ denote the diagonal elements of $\Sigma_{k,r}$
	\end{tablenotes}	  
\end{threeparttable}
\end{table}%

\newpage
\begin{table}[htbp]
  \centering
  \caption{Root mean square error of parameter estimates for model with $K=4$ under random effects with $t$ distribution} 
        \begin{threeparttable} 
        \begin{tabular}{lcccccccccccccccc}
    \toprule
                      & \multicolumn{4}{c}{$\bm{\alpha} =  (0.25, 0.25, 0.25) $}                      & \multicolumn{4}{c}{$\bm{\alpha}  =  (0.8, 0.8, 0.8) $}                        & \multicolumn{4}{c}{$\bm{\alpha} =  (1, 1, 1) $}                               & \multicolumn{4}{c}{$\bm{\alpha} =  (1, 0.25, 0.8) $} \\
    \midrule
                      & $k=1$             & $k=2$             & $k=3$             & $k=4$             & $k=1$             & $k=2$             & $k=3$             & $k=4$             & $k=1$             & $k=2$             & $k=3$             & $k=4$             & $k=1$             & $k=2$             & $k=3$             & $k=4$ \\
    \midrule
    $\pi_k$           & 0.123             & 0.158             & 0.105             & 0.266             & 0.035             & 0.042             & 0.016             & 0.056             & 0.003             & 0.005             & 0.003             & 0.006             & 0.013             & 0.011             & 0.009             & 0.009 \\
    $r = 1$ (Continous) &                   &                   &                   &                   &                   &                   &                   &                   &                   &                   &                   &                   &                   &                   &                   &  \\
    $\alpha_r$        & \multicolumn{4}{c}{0.047}                                                     & \multicolumn{4}{c}{0.080}                                                     & \multicolumn{4}{c}{0.011}                                                     & \multicolumn{4}{c}{0.150} \\
    $\gamma_{1k,r}$   & 0.063             & 0.060             & 0.051             & 0.059             & 0.064             & 0.075             & 0.062             & 0.057             & 0.063             & 0.070             & 0.056             & 0.062             & 0.058             & 0.059             & 0.053             & 0.055 \\
    $\gamma_{2k,r}$   & 0.010             & 0.013             & 0.011             & 0.010             & 0.012             & 0.013             & 0.011             & 0.011             & 0.011             & 0.011             & 0.010             & 0.012             & 0.012             & 0.011             & 0.011             & 0.012 \\
    $\gamma_{3k,r}$   & 0.000             & 0.000             & 0.000             & 0.000             & 0.000             & 0.000             & 0.000             & 0.000             & 0.000             & 0.000             & 0.000             & 0.000             & 0.000             & 0.000             & 0.000             & 0.000 \\
    $\Sigma_{11k,r}$  & 0.000             & 0.000             & 0.000             & 0.000             & 0.000             & 0.000             & 0.000             & 0.000             & 0.000             & 0.000             & 0.000             & 0.000             & 0.000             & 0.000             & 0.000             & 0.000 \\
    $\Sigma_{22k,r}$  & 0.000             & 0.000             & 0.000             & 0.000             & 0.000             & 0.000             & 0.000             & 0.000             & 0.000             & 0.000             & 0.000             & 0.000             & 0.000             & 0.000             & 0.000             & 0.000 \\
    $r = 2$  (Count)  &                   &                   &                   &                   &                   &                   &                   &                   &                   &                   &                   &                   &                   &                   &                   &  \\
    $\alpha_r$        & \multicolumn{4}{c}{0.269}                                                     & \multicolumn{4}{c}{0.169}                                                     & \multicolumn{4}{c}{0.180}                                                     & \multicolumn{4}{c}{0.028} \\
    $\gamma_{1k,r}$   & 0.524             & 0.766             & 0.744             & 0.671             & 0.529             & 0.763             & 0.731             & 0.682             & 0.534             & 0.760             & 0.723             & 0.658             & 0.527             & 0.769             & 0.725             & 0.669 \\
    $\gamma_{2k,r}$   & 0.011             & 0.048             & 0.053             & 0.013             & 0.011             & 0.048             & 0.051             & 0.012             & 0.012             & 0.048             & 0.052             & 0.013             & 0.011             & 0.048             & 0.052             & 0.013 \\
    $\Sigma_{11k,r}$  & 0.011             & 0.011             & 0.004             & 0.001             & 0.011             & 0.011             & 0.004             & 0.001             & 0.011             & 0.011             & 0.004             & 0.001             & 0.011             & 0.011             & 0.004             & 0.001 \\
    $\Sigma_{22k,r}$  & 0.000             & 0.000             & 0.000             & 0.000             & 0.000             & 0.000             & 0.000             & 0.000             & 0.000             & 0.000             & 0.000             & 0.000             & 0.000             & 0.000             & 0.000             & 0.000 \\
    $r = 3$ (Binary)  &                   &                   &                   &                   &                   &                   &                   &                   &                   &                   &                   &                   &                   &                   &                   &  \\
    $\alpha_r$        & \multicolumn{4}{c}{0.236}                                                     & \multicolumn{4}{c}{0.079}                                                     & \multicolumn{4}{c}{0.017}                                                     & \multicolumn{4}{c}{0.108} \\
    $\gamma_{1k,r}$   & 0.781             & 7.212             & 1.402             & 0.748             & 0.347             & 4.071             & 0.288             & 0.299             & 0.207             & 0.343             & 0.179             & 0.211             & 0.289             & 2.655             & 0.210             & 0.257 \\
    $\gamma_{2k,r}$   & 0.198             & 4.911             & 1.380             & 1.407             & 0.017             & 3.110             & 0.018             & 0.018             & 0.011             & 0.024             & 0.010             & 0.013             & 0.015             & 1.178             & 0.014             & 0.015 \\
    $\Sigma_{11k,r}$  & 0.000             & 0.000             & 0.000             & 0.000             & 0.000             & 0.000             & 0.000             & 0.000             & 0.000             & 0.000             & 0.000             & 0.000             & 0.000             & 0.000             & 0.000             & 0.000 \\
\cmidrule{1-17}    
    \end{tabular}    
    \begin{tablenotes} 
       \item $\Sigma_{11k,r}$ and $\Sigma_{22k,r}$ denote the diagonal elements of $\Sigma_{k,r}$
	\end{tablenotes}	  
\end{threeparttable}
\end{table}%

\newpage
\KOMAoptions{paper=portrait,pagesize}